\documentclass{aa}  

\usepackage{graphicx}
\usepackage{amsmath}	
\usepackage{amssymb}	
\usepackage{multicol}   
\usepackage{bm}		
\usepackage{pdflscape}	
\usepackage{bm}		
\usepackage{threeparttable,lscape}
\usepackage{natbib}
\usepackage{rotating}
\usepackage{color}
\usepackage{xcolor}
\usepackage{soul}
\usepackage{appendix}
\usepackage{multirow,array}
\usepackage{longtable}
\usepackage{hyperref}
\usepackage{txfonts}
\usepackage[T1]{fontenc}
\usepackage{ae,aecompl}
\usepackage{ulem}

\newcommand{\kms}{\,km\,s$^{-1}$} 
\newcommand{\naid}{\ion{Na}{i}\,D\,}
\newcommand{\naids}{\ion{Na}{i}\,D}
\newcommand{\caii}{\ion{Ca}{ii}\,}
\newcommand{\ki}{\ion{K}{i}\,}
\newcommand{\sii}{\ion{Si}{ii}\,}

\usepackage{hyperref}
\definecolor{yaleblue}{rgb}{0.1,0.3,0.9}
\definecolor{lava}{rgb}{0.81, 0.06, 0.13}
\definecolor{forestgreen}{rgb}{0.0, 0.45, 0.13}
\hypersetup{colorlinks=true, linkcolor=lava, urlcolor=forestgreen, citecolor=yaleblue}

\newcommand{\orcid}[1]{\href{https://orcid.org/#1}{\includegraphics[width=10pt]{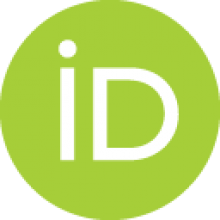}}}

\begin{document}

   \title{Narrow absorption lines from intervening material in supernovae}

   \subtitle{I. Measurements and temporal evolution}

   \author{Santiago Gonz\'alez-Gait\'an
          \inst{1}\orcid{0000-0001-9541-0317}
          \and
          Claudia P. Guti\'errez
          \inst{2,3}\orcid{0000-0003-2375-2064}
          \and
          Joseph P. Anderson
          \inst{4,5}\orcid{0000-0003-0227-3451}
          \and
          Antonia Morales-Garoffolo
          \inst{6}
          \and
          Lluis Galbany
          \inst{3,2}\orcid{0000-0002-1296-6887}
          \and
          Sabyasashi Goswami
          \inst{1,6}
          \and
          Ana M. Mour\~ao
          \inst{1}\orcid{0000-0002-0855-1849}
          \and 
          Seppo Mattila
          \inst{7,8}\orcid{0000-0001-7497-2994}
          \and 
          Mark Sullivan
          \inst{9}\orcid{0000-0001-9053-4820}
          }

   \institute{CENTRA, Instituto Superior T\'ecnico, Universidade de Lisboa, Av. Rovisco Pais 1, 1049-001 Lisboa, Portugal\\
    \email{gongsale@gmail.com}
    \and
    Institut d’Estudis Espacials de Catalunya (IEEC), Gran Capit\`a, 2-4, Edifici Nexus, Desp. 201, E-08034 Barcelona, Spain\\
    \email{cgutierrez@ice.csic.es}
    \and
    Institute of Space Sciences (ICE, CSIC), Campus UAB, Carrer de Can Magrans, s/n, E-08193 Barcelona, Spain
    \and
    European Southern Observatory, Alonso de C\'ordova 3107, Casilla 19, Santiago, Chile
    \and
    Millennium Institute of Astrophysics MAS, Nuncio Monsenor Sotero Sanz 100, Off. 104, Providencia, Santiago, Chile
    \and
    Department of Applied Physics, School of Engineering, University of C\'adiz, Campus of Puerto Real, E-11519 C\'adiz, Spain
    \and
    Tuorla Observatory, Department of Physics and Astronomy, University of Turku, FI-20014 Turku, Finland
    \and
    School of Sciences, European University Cyprus, Diogenes Street, Engomi, 1516 Nicosia, Cyprus
    \and
    School of Physics and Astronomy, University of Southampton, Southampton, SO17 1BJ, UK
   }
   \date{}

 
\abstract
{Narrow absorption features in nearby supernova (SN) spectra are a powerful diagnostic of the slow-moving material in the line of sight: they are extensively used to infer dust extinction from the host galaxies, and they can also serve in the detection of circumstellar material originating from the SN progenitor and present in the vicinity of the explosion. Despite their wide use, very few studies have examined the biases of the methods to characterize narrow lines, and not many statistical analyses exist. This is the first paper of a series in which we present a statistical analysis of narrow lines of SN spectra of various resolutions. We develop a robust automated methodology to measure the equivalent width (EW) and velocity of narrow absorption lines from intervening material in the line of sight of SNe, including \naid, \caii\,H\&K, \ki and diffuse interstellar bands (DIBs). We carefully study systematic biases in heterogeneous spectra from the literature by simulating different signal-to-noise, spectral resolution, slit size and orientation and present the real capabilities and limitations of using low- and mid-resolution spectra to study these lines. In particular, we find that the measurement of the equivalent width of the narrow lines in low-resolution spectra is highly affected by the evolving broad P-Cygni profiles of the SN ejecta, both for core-collapse and type Ia SNe, inducing a conspicuous apparent evolution. Such pervading non-physical evolution of narrow lines might lead to wrong conclusions on the line-of-sight material, e.g. concerning circumstellar material ejected from the SN progenitors. We present thus an easy way to detect and exclude those cases to obtain more robust and reliable measurements. Finally, after considering all possible effects, we analyse the temporal evolution of the narrow features in a large sample of nearby SNe to detect any possible variation in their EWs over time. We find no time evolution of the narrow line features in our large sample for all SN types.}

\authorrunning{Gonz\'alez-Gait\'an, 
Guti\'errez et al.}
\titlerunning{Narrow absorption lines in SNe}
\maketitle
%

\section{Introduction}

The light emitted by astrophysical sources detected from Earth is subject to interaction with matter that lies along the line of sight between the source and the observer. Studying the effects of this line of sight material on the background light source gives information on the nature of that material, whether it is close to the light source, e.g. circumstellar material (CSM), within galaxies, i.e. the interstellar medium (ISM), or between galaxies, i.e. the intergalactic medium (IGM). It is important to understand these processes: CSM for stellar mass loss and ISM and IGM for galaxy enrichment and evolution. At the same time, this intervening material is also extremely important for understanding the light sources themselves as it absorbs the light, especially dimming it at optical wavelengths and re-emitting it at longer wavelengths. It thus impacts on our understanding of the brightness and properties of the emitters. 

This material in the line of sight may be composed of dust and gas. Dust grains are a product of stellar evolution and responsible for the effect of absorption and reddening of optical starlight \citep[e.g.,][]{Draine11}. Molecular gas is known to exist, e.g. in the form of hydrogen (H$_2$) and of carbon monoxide (CO) \citep{Dame01} but also in the form of diffuse interstellar bands (DIBs) which trace as-yet-unidentified molecules \citep[e.g.][]{Lan15}. On the other hand, the atomic gas phase component of the intervening material is mostly composed of hydrogen. Nevertheless, a fraction of metals is also present and provides valuable information on the chemistry, ionization state and gas temperature of the line-of-sight matter. These different species will absorb the light at characteristic wavelengths, generating detectable absorption lines in the spectroscopic observations.

Optical interstellar absorption lines in the form of sodium, calcium, potassium and DIBs have long been observed in spectroscopic binaries \citep[e.g.,][]{Heger19, Heger19a, Young22}, in early-type stars \citep{Struve28, Wilson37, Merrill37, Sanford37, Evans41}, in novae \citep{Merrill35}, in galaxies \citep[e.g.][]{Heckman00}, in active galactic nuclei \citep[e.g.][]{Baron16} and in supernovae (SN; e.g. \citealt{Penston80}). In contrast to material moving at high velocity, as is the case for SN ejecta, these intervening lines will appear much narrower than the characteristic P-Cygni profiles of fast-expanding material. Today, narrow interstellar lines are routinely detected in the optical spectra of a great number of astrophysical transients \citep[e.g.,][]{Galazutdinov05, Megier09, Chen10, Park15}. The relation of the strength of these lines with colour excess and extinction has been studied in depth \citep[e.g.][]{Merrill38, Spitzer48, Buscombe68, Hobbs74} providing empirical relations between the amount of dust extinction and strength of the gas lines \citep[e.g.,][]{Richmond94, Munari97, Turatto03, Poznanski12, Murga15}, with caveats for the sodium line since it saturates \citep{Munari97, Poznanski11}. Alternative lines like potassium \citep[e.g.][]{Hobbs74, Munari97, Galazutdinov05, Phillips13} and DIBs \citep[][]{Kos13, Phillips13, Krelowski19} have been claimed to be superior extinction tracers. 

SNe offer a particularly good laboratory to investigate the material in the line of sight as they are bright and are found in many different environments. On the other hand, understanding this material is also key to understanding SN progenitor systems and their explosions, as it can give us information on stellar evolution and mass loss. Indeed, the study of narrow line-of-sight lines in transient objects has increased during the last few years, partly due to the detection of time-variable absorption features in high-resolution spectra. While intermediate-luminosity red transients embedded in dusty cocoons show clear signs of sodium line strength evolution \citep{Byrne23}, the only SN coming from the core collapse of a massive star showing this evolution is the broad-lined type Ic, SN~2012ap \citep{Milisavljevic13}, with changes in the strength of the DIBs but no variations in the sodium lines. The authors conclude that the material producing these variations is nearby, and the SN interacts with it. More surprisingly, a handful of type Ia SNe -- the thermonuclear runaway of white dwarfs in binary systems -- have shown evolving sodium lines in high-resolution spectra: SN~2006X \citep{Patat07}, SN~1999cl \citep{Blondin09}, SN~2007le \citep{Simon09}, SN~2013gh \citep{Ferretti16}, and SN~2014J with time-varying potassium instead of sodium \citep{Graham15}. This has significant implications for the debate surrounding the SN progenitor system: it has been suggested that such variable lines could originate from the CSM ejected by the progenitor system, in principle favouring the single degenerate scenario of a binary system composed of a white dwarf and a main-sequence or red giant \citep[e.g.][]{Moore12}. However, double degenerate models composed of two white dwarfs could potentially also exhibit CSM characteristics \citep{Raskin13, Shen13, Levanon19}. 
Such CSM claims are further emphasized with the observation of a statistical excess of blueshifted (vs redshifted) sodium absorption lines in high-resolution SN~Ia spectra \citep{Sternberg13, Maguire13, Phillips13, Clark21}, that may also be related to other SN properties such as ejecta velocity and SN colour \citep{Foley12}. 

On the other hand, measurements of narrow lines from intervening material in SN~Ia spectra of lower resolution have also shown intriguing correlations between line strength and intrinsic SN properties such as the nebular velocity shifts \citep{Forster12} and the late colour decline rate \citep{Forster13}, arguing again for a possible interaction of nearby material with a sub-group of SNe~Ia. This has been recently confirmed by \citet{Wang19} also with low/mid-resolution spectra who additionally find possible line evolution in the interacting SNe~Ia. 

A significant caveat of the above studies is that lower resolution and signal-to-noise ratio (S/N) spectra may affect or even prevent proper measurements. Can low and mid-resolution spectra be used to infer narrow line equivalent widths and velocities? Can weaker lines such as DIBs and potassium reliably be detected in such spectra? Can the presence of the high-velocity, broad features from the SN ejecta interfere with the narrow line measurements? Are there any biases that could simulate an apparent evolution or blueshift? Given the large amount of SN spectra gathered in the literature and the scarcity of methods to investigate the material in the light of sight of the SN, the capability to robustly use these spectra for narrow lines studies could provide powerful, statistically meaningful results on the presence and distribution of these species in the intervening matter. 

In order to investigate narrow lines, we use here simulations and a large sample of observed SNe from the literature that consists of more than 1000 objects of different types --including SNe~Ia, SNe~II (classical plateau and linearly declining hydrogen-rich SNe), stripped-envelope SNe (SESNe) with SNe~Ic, SNe~Ib and SNe~IIb, and interacting SNe including SNe~IIn, SNe~Ibn and SNe~Icn. The spectra and simulations include various spectral resolutions that allow us to test several different possible biases in the measurements of equivalent widths and velocities of narrow line-of-sight lines. After defining the limitations of our technique, we also look into the likelihood that these lines in SN spectra evolve over time. In forthcoming papers, we will investigate the differences in these lines for various SN types, SN properties and environmental characteristics.

The paper is organised as follows. Section~\ref{sec:sodiim} introduces the lines from intervening matter analysed in this study. We present the data sample in Section~\ref{sec:sample} and the automated line measurement in Section~\ref{sec:measurements}. The systematic biases are presented in Section~\ref{sec:bias}, while the evolution analysis and discussion are in Section~\ref{sec:evolution}. We conclude in Section~\ref{sec:conclusions}.

\section{Line-of-sight absorption lines}
\label{sec:sodiim}

The properties of the ISM of a SN host galaxy and,  more specifically, the environment where the SN exploded -- or of the CSM ejected by the progenitor prior to explosion -- can be studied with spectroscopy. The spectral resolution, usually given by $R=\lambda/\Delta\lambda$ reveals the capacity of the instrument to resolve narrower lines; we define here high-resolution spectra as those with $R\gtrsim10000$ (or $\Delta \mathrm{v}<30$km/s; \citealt{Appenzeller86}) and mid-resolution as $5000<R<1000$ (or 30km/s$<\Delta \mathrm{v}<$60km/s). In nearby objects exploding in reddened environments, it is possible to detect narrow lines close to the host galaxy's rest-frame line wavelength, but also due to the MW close to the observer-frame line wavelength, at a variety of spectral resolutions and signal-to-noise ratios, as will be shown later (sections~\ref{sec:specres} and~\ref{sec:signois}).

We outline here the main optical absorption lines that can be measured and utilised to investigate the ISM and the extinction towards SNe. 

\begin{itemize}
\item Sodium (\naid) lines are very prominent and the typical tracers of the ISM. They were discovered in the solar spectrum by Fraunhofer in 1814. They are the doublet from the fine structure splitting of the excitation states of neutral sodium at $\lambda=5891.58$\,\r{A} for D$_2$ and $\lambda=5897.56$\,\r{A} for D$_1$. Neutral sodium has an ionization potential of 5.14 eV \citep{Morton03}.
\item Calcium (\caii H\&K) lines were also seen by Fraunhofer in sunlight and, similarly, they correspond to the doublet of the fine structure splitting of the excitation states of the singly ionized calcium at $\lambda=3934.8$\,\r{A} for K and $\lambda=3969.6$\,\r{A} for H, with an ionization potential of 11.87 eV. 
\item Potassium (\ki) lines were first detected by \citet{Dunham37}. This resonant doublet from the fine structure splitting of the neutral potassium excitation states occurs at $\lambda=7664.90$\,\r{A} for K$_1$ and $\lambda=7698.97$\,\r{A} for K$_2$. Neutral potassium has an ionization potential of 4.34 eV.
\item The diffuse interstellar bands (DIBs) were first discovered in stellar spectra by \citet{Heger1922}. Today, hundreds of DIBs from unidentified molecules in the ISM have been discovered (see \citealt{Herbig95}). Proposed carriers include polycyclic aromatic hydrocarbons (PAHs), fullerenes and other hydrocarbons ions and molecules \citep{Salama99, Kroto85, Motylewski00}. We choose here three of the strongest and more isolated DIBs at $\lambda= 4428.2, 5780.5, 6283.8$\,\r{A}.
\end{itemize}

Table~\ref{tab:lines} summarizes these narrow lines and their characteristics. We also include the "integration window", a velocity interval around the central wavelength in which the line will be measured (see section~\ref{sec:ew_meas}). It is related to the strength of the line and therefore the wavelength range it may cover.

\begin{table}
\centering
\caption{Narrow spectral features from the intervening medium considered in this study.}
\label{tab:lines}
\begin{tabular}{m{1.4cm}m{1.8cm}m{1.6cm}m{1.8cm}}
\hline
Line & Wavelength (\AA) & Ionization potential (eV) & Integration window (km/s)\\
\hline
\naid & 5889.95 5895.92 & 5.14 & $\pm$1000\\
\caii H & 3969.60 & 11.87 & $\pm$900\\
\caii K & 3934.80 & 11.87 & $\pm$900\\
\ki 1 & 7664.90  & 4.34 & $\pm$600\\
\ki 2 & 7698.97  & 4.34 & $\pm$600\\
DIB 4428 & 4428.20 & -- & $\pm$750\\
DIB 5780 & 5780.50 & -- & $\pm$600\\
DIB 6283 & 6283.80 & -- & $\pm$750\\
\hline
\end{tabular}
\end{table}

\begin{figure*}
\centering
\includegraphics[width=\textwidth]{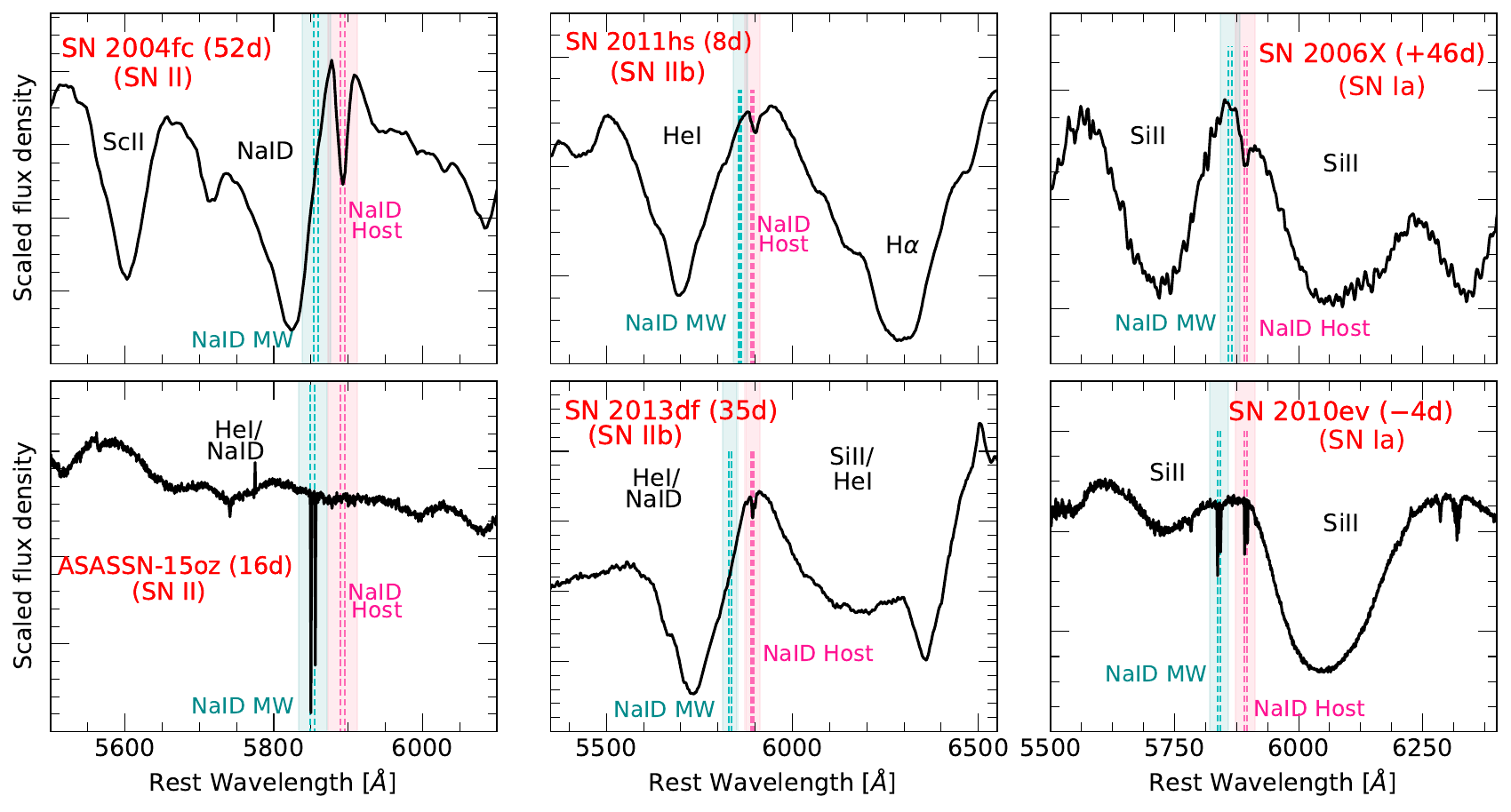}
\caption{Location of the narrow \naid absorption lines in SNe~II (left), SESNe (middle) and SNe~Ia (right panel). The names and phases with respect to the $B$-band maximum in their observed light curves are marked in each inset. Low-resolution spectra are presented in the top panels (pseudo-resolution -- see Section~\ref{sec:sample} -- of $R_p=$ 1951,3087 and 3869 from left to right), while higher-resolution spectra ($R_p=$ 11877, 9306 and 14879 from left to right) are at the bottom. The vertical dashed lines indicate the position of the \naid for the MW (cyan) and the host galaxy (magenta), with the integration window of $\pm$ 1000km/s shown in shaded areas. 
}
\label{naidSN}
\end{figure*}

SNe in low-reddening environments may only display the strongest line-of-sight lines, which normally are the \naid absorption features. If the strength of the absorption line is very low, it indicates negligible material in the line of sight and very low reddening. Figure~\ref{naidSN} shows the location of the narrow \naid features for some low-redshift ($z<0.1$) SNe observed with a low (top panels) and high (bottom panels) resolution instrument. In SNe~II (left panels), the narrow \naid lines are on the top of the \ion{He}{i} $\lambda5876$ broad line from the ejecta at early phases, but after 30-40 days, they are around the peak of the broad \naid emission line (from the SN). In SESNe (middle panels), they are between the \ion{He}{i}/\naid and the \ion{Si}{ii}/\ion{He}{i} or H$\alpha$ (in SNe~IIb), while in SNe~Ia (right panels), the narrow \naid features are between the \ion{Si}{ii} lines. The effect these broad lines have in the narrow lines will be further discussed in Section~\ref{sec:bias}.

In high-resolution spectra, the narrow \naid lines can be resolved. It is possible to detect four lines in the spectra (see, for example, bottom panel of Figure~\ref{naidSN}), two from the MW ($\lambda5890$ and $\lambda5896$) and two from the host galaxy ($\lambda5890 \times (1+z)$ and $\lambda5896 \times (1+z)$, where $z$ is the SN redshift), as long as a single cloud component is present. For low-resolution spectra, only two lines can be detected (one that corresponds to the MW ($\lambda5893$, on average) and another related to the host galaxy ($\lambda5893\times (1+z)$).
When the SN is very nearby ($z<0.04$; e.g. SN~1987A), the MW and the host galaxy lines can be blended, and just a single narrow absorption line is visible. These cases are not included in the current analysis (see the selection criteria in Sections~\ref{sec:sample} and~\ref{sec:bias}). 

The narrow line-of-sight lines play an important role in estimating the reddening in the line of sight of SNe. Theoretically, this is somewhat expected as the line equivalent width (EW) of a given gas species $X$ is given by:  

\begin{equation}
\mathrm{EW_X} = \int \left(1-e^{-\tau_X(\lambda)}\right)\,d\lambda,
\end{equation}
where $\tau_X$ is the optical depth, which is a function of the column density of the species, $N_X$, multiplied by the absorption coefficient, $\kappa_0$. In turn, the gas column density depends on the gas-to-dust ratio, $f_{\mathrm{gd}}$, the mass fraction of the gas, $f_X$, and the ionization fraction, $f_{\mathrm{ion}}$. For low optical depths, EW$_X \propto \tau_X \propto N_X$, so that:

\begin{align}
    \mathrm{EW_X} \propto \tau_X =
    \kappa_0N_X =
    \kappa_0(1-f_{\mathrm{ion}})f_{X}f_{\mathrm{gd}}N_{\mathrm{dust}},
\end{align}
where the dust column density, $N_{\mathrm{dust}}$, is directly proportional to the extinction $A_V$:
\begin{align}\label{eq:EW-Av}
    \mathrm{EW_X} \propto \kappa_0(1-f_{\mathrm{ion}})f_{X}f_{\mathrm{gd}}A_{V}
\end{align}

As the optical depth increases, the line profile becomes saturated, and the equivalent width does not increase linearly with column density. Indeed, the strength at the central wavelength stalls whereas the Doppler broadening increases, so that EW$_X \propto \sqrt{\ln{ N_X}}$. At even larger optical depths, the collisional broadening dominates and EW$_X \propto \sqrt{N_X}$.

The linear relation between EW and $A_V$ at low optical depths (eq.~\ref{eq:EW-Av}) depends strongly on other factors that are quite uncertain and vary substantially in our MW. For example, the low ionization potentials of the gaseous metals (below the Hydrogen ionization potential of 13.6 eV, see Table~\ref{tab:lines}) mean that they are easily photo-ionized, changing the ionization fraction. Multiple gas clouds in the line of sight can also create several spectral components whose varying velocities cannot be disentangled in mid- and low-resolution spectra and thus contribute to the saturation of the line.   

Despite these uncertainties in the theoretical relation,  it has been shown empirically, as mentioned above, that the strength of the narrow \naid intervening lines correlates with dust extinction, based either on nearby stars  \citep[e.g.,][]{Munari97}, SNe \citep{Turatto03,Sollerman05} or large samples of extragalactic sources whose light is absorbed in the MW gas and dust \citep{Poznanski12, Murga15}. The trends are limited by the large dispersion and a distinct saturation for EW values larger than 0.5 \AA\ in low-resolution data \citep{Munari97, Poznanski11}, and 1.8 \AA\ in higher resolution spectra (1 \AA\ for \naid$_2$ and 0.8 \AA\ for \naid$_1$; \citealt{Poznanski12}) respectively. \citet{Poznanski11} suggested that the EW estimated from low-resolution spectra is a bad estimator of $E(B-V)$, while \citet{Phillips13} argued that
the \naid gives an overabundance of $E(B-V)$ for several SNe~Ia and the DIB at 5780 \AA\ is a better tracer of reddening than \naid. Indeed, \cite{Baron15} demonstrate that eight different DIBs correlate well with extinction for a large sample of extragalactic objects from the Sloan Digital Sky Survey.

In the context of SNe, the presence of the high-velocity, wide features from the SN makes the measurement of the EW difficult (see section~\ref{sec:broadline}). Additionally, time-variable \naid absorption lines detected in a few SNe~Ia \citep{Patat07, Blondin09, Simon09} also complicate their use as reddening tracers. However, considering their easy identification and the scarcity of other methods to constrain dust extinction along the line of sight, their use is still widespread. That is why a robust methodology and a careful study of systematic biases, as the one presented here, are needed.

\section{Sample}
\label{sec:sample}

\begin{figure*}
\centering
\includegraphics[width=\textwidth]{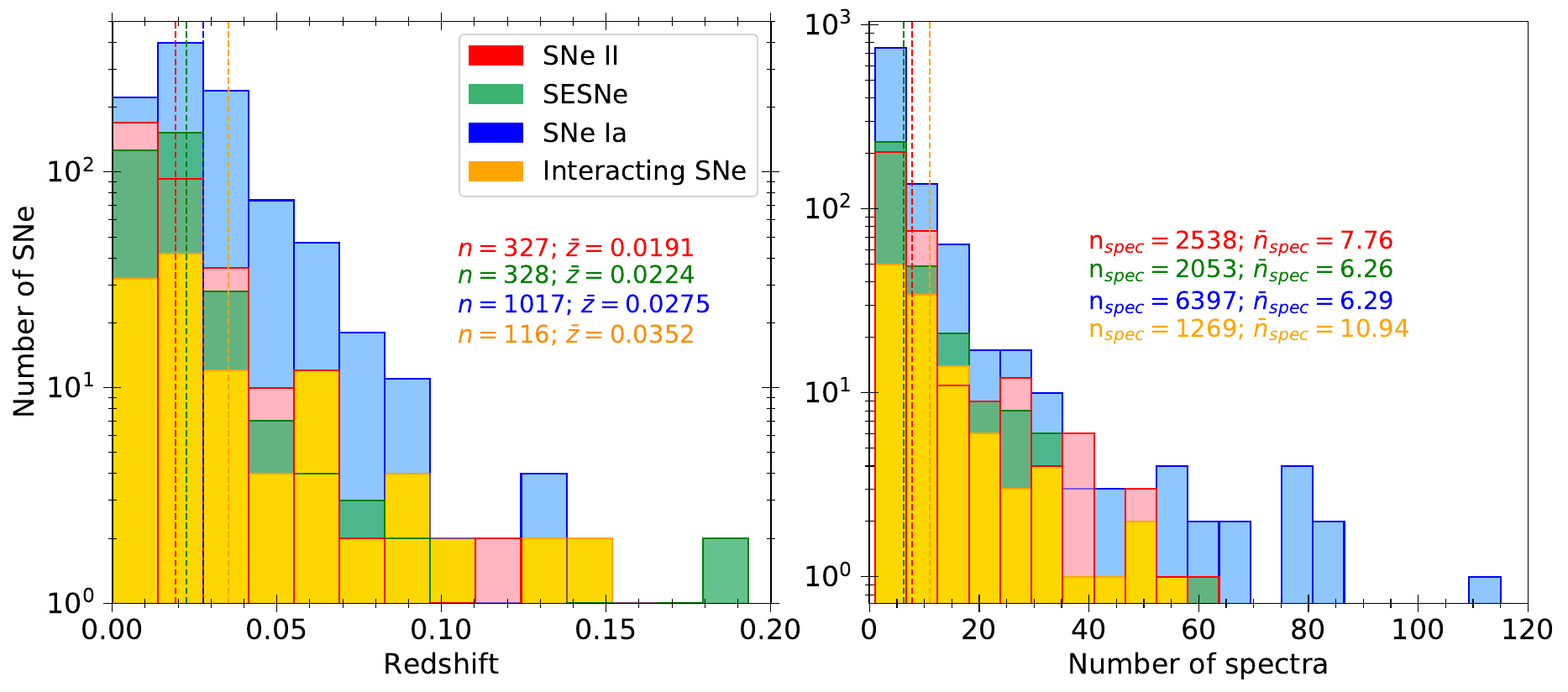}
\caption{\textbf{Left:} Distribution of heliocentric redshifts for the 1719 SNe included in our sample. \textbf{Right:} Histogram of the number of spectra per SN.
}
\label{histograms}
\end{figure*}

The data analysed in this paper consists of a large heterogeneous sample from the literature obtained through large web compilations such as SUSPECT \citep{Richardson01}, WISEREP\footnote{\url{https://www.wiserep.org/}} \citep{Yaron12}, the Open Supernova Catalog\footnote{\url{https://sne.space/}} \citep{Guillochon17}, and from individual journal publications. It contains SNe~Ia from the thermonuclear runaway of white dwarfs in binary systems, including the peculiar 91bg-like, 91T-like, 02es-like and Iax; it contains hydrogen-rich SNe (SN~II) from the core-collapse of massive stars that have retained their hydrogen envelopes, including both old light-curve classifications, SNe~IIP (plateau) and SNe~IIL (fast-decline); it contains stripped-envelope (SESNe) from the core-collapse of massive stars that have lost their outer layers: IIb with hydrogen shown only initially in their spectra but not later on, Ib without hydrogen throughout their evolution, Ic without hydrogen nor helium in their spectra and Ic-BL with broad lines from a central engine; and it contains various transients related to stellar outbursts and death of massive stars with clear signs of interaction with circumstellar material like SNe~IIn, SNe~Ibn, SNe~Icn, and SN impostors, collectively called here "SNe-int".

This dataset consists of 12257 spectra of 1788 SNe observed between 1937 and 2023. 2538 of these spectra correspond to 327 SNe~II, 2053 spectra to 328 SESNe, 1269 spectra to 116 SNe-int and 6397 spectra to 1017 SNe~Ia. The sample also includes 174 spectra with mid- and high-resolution. Most of them are from SNe~Ia. Only 19 are from SESNe, and 60 are from SNe~II. All spectra are corrected for MW extinction with the maps of \citet{Schlafly11} and brought to the rest frame. Due to the heterogeneous nature of the sources, this sample is not complete, neither in volume nor in magnitude. The sample details can be found in Table~\ref{table_SNe} in the Appendix.

Figure~\ref{histograms} (left panel) shows the redshift distribution of our sample. The mean redshift value is 0.026, while the median is 0.019. The nearest object is SN~1987A with a redshift of 0.00001, while the farthest object is iPTF-13ajg with a redshift of 0.74. 
In the right panel of Figure~\ref{histograms}, we present the distribution of spectra per object. 720 SNe have at least two spectra, 873 SNe have at least three spectra, and 937 SNe have between 3 and 10 spectra. The objects with the most data are the type Ia SN~2005cf (115), SN~2006X (85) and SN~2001V (82). On average, we have 6.7 spectra per SN. For the sample of SNe~II, the average is 7.6, 6.4 for SESNe, and 6.4 for SNe~Ia.

\begin{figure}
\centering
\includegraphics[width=\columnwidth]{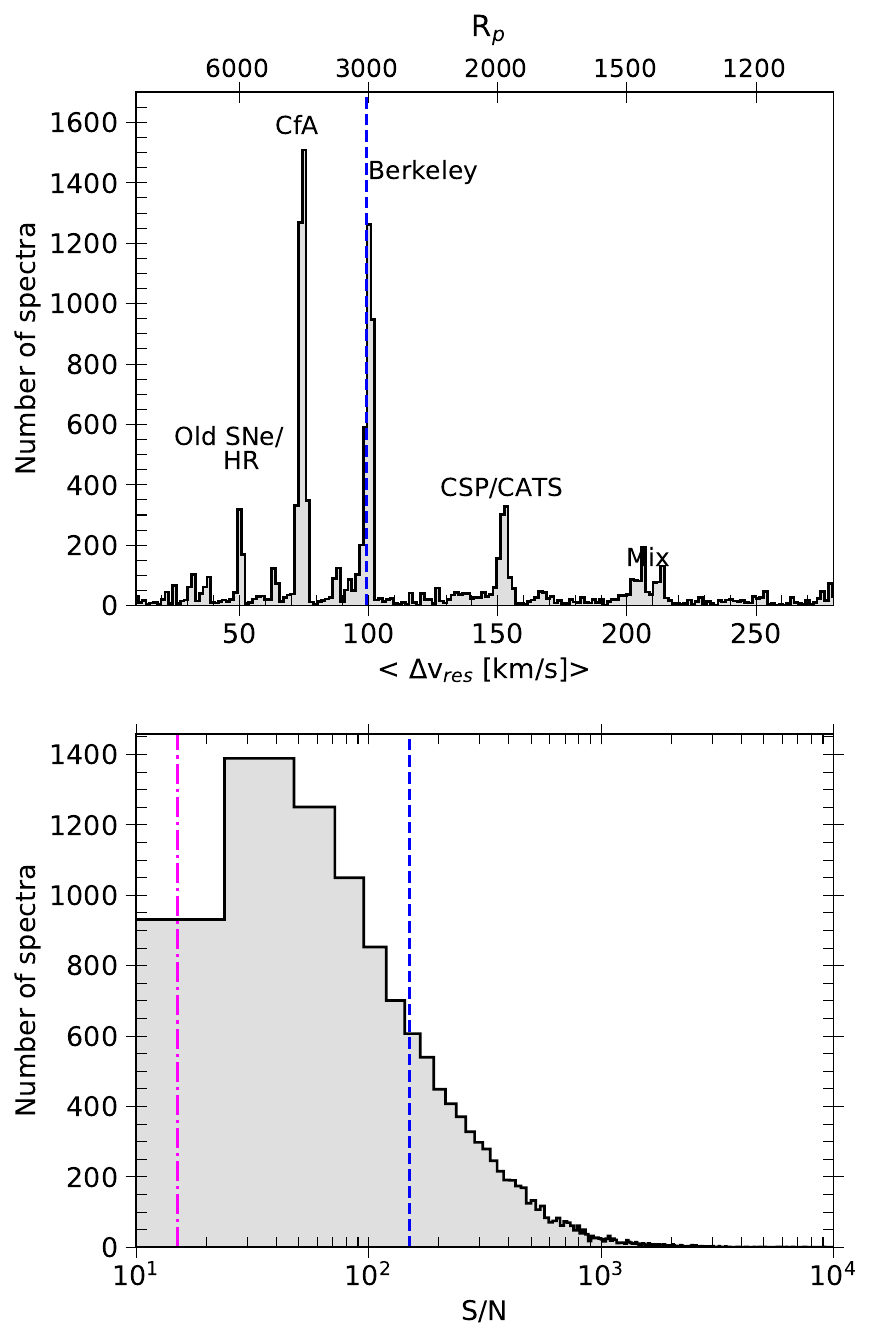}
\caption{Distribution of average velocity resolution (\textbf{top})  around the sodium doublet and signal-to-noise (\textbf{bottom}) for all spectra in our sample. Vertical blue dashed lines show the median values of the distributions ($\Delta \mathrm{v}_{\mathrm{res}}=99.2$ km/s and $S/N=155$).}
\label{histograms_sigres}
\end{figure}

To give a better idea of the spectral quality of this large and diverse sample, we show in the top panel of Figure~\ref{histograms_sigres} the average resolution of all spectra, where we define this resolution as the average sampling around 10,000~\kms of the narrow line (in this case \naid): $\Delta \mathrm{v}_{\mathrm{res}} = \Delta\lambda_p c/\lambda_{\mathrm{rest}}$ and we also show the corresponding pseudo-resolution, $R_p=\lambda_{\mathrm{rest}}/\Delta\lambda_P$, in the upper axis. $\Delta\lambda_p$ is the wavelength sampling provided in the spectral files instead of the full-width half-maximum of unresolved lines, which characterizes an instrument resolution. This is why we call $R_p$ a pseudo-resolution, which is only a proxy for the real spectral resolution of the instrument. 
In particular, some spectra have higher wavelength spacing without meaning the spectral resolution is better. This is especially true for many old SN spectra from years before 1995 that have a narrower spacing data format but are not of particularly high resolution. These represent most of the first peak shown in the distribution, although some real high-resolution spectra are also present. Other peaks in the distribution are mostly but not exclusively populated by the spectra of large surveys with its instruments and their characteristic wavelength spacing: the supernova group at the Harvard-Smithsonian Center for Astrophysics\footnote{\url{https://lweb.cfa.harvard.edu/supernova/SNarchive.html}} \citep[CfA; e.g.][]{Blond12, Modjaz14, Hicken17}, the UC Berkeley SN group\footnote{\url{https://heracles.astro.berkeley.edu/sndb/}} \citep[Berkeley; e.g.][]{Silverman12, Shivvers19, Stahl20} and Carnegie Supernova Project\footnote{\url{https://csp.obs.carnegiescience.edu/data}} \citep[CSP; e.g.][]{Folatelli13, Gutierrez17a, Stritzinger23} are the most characteristic ones. 

On the bottom panel of Figure~\ref{histograms_sigres}, we also show the average S/N in the same wavelength regime calculated with the inverse of the root mean square (RMS) of the difference between the original spectrum and a smoothed spectrum with a cosine kernel. The distribution is exponentially falling with a mode around 30 and a median of 150. 

With the aim of doing a systematic analysis and decreasing the bias in our sample, we will define several criteria for our data selection (see Table~\ref{tab:cuts}) that will be discussed in the next sections.

\section{Automated line measurement}
\label{sec:measurements}

In this section, we outline the automated methodology we have developed to measure EWs and velocities in a diverse set of spectra.

\subsection{Equivalent width measurement} 
\label{sec:ew_meas}

\begin{figure}
\centering
\includegraphics[width=\columnwidth]{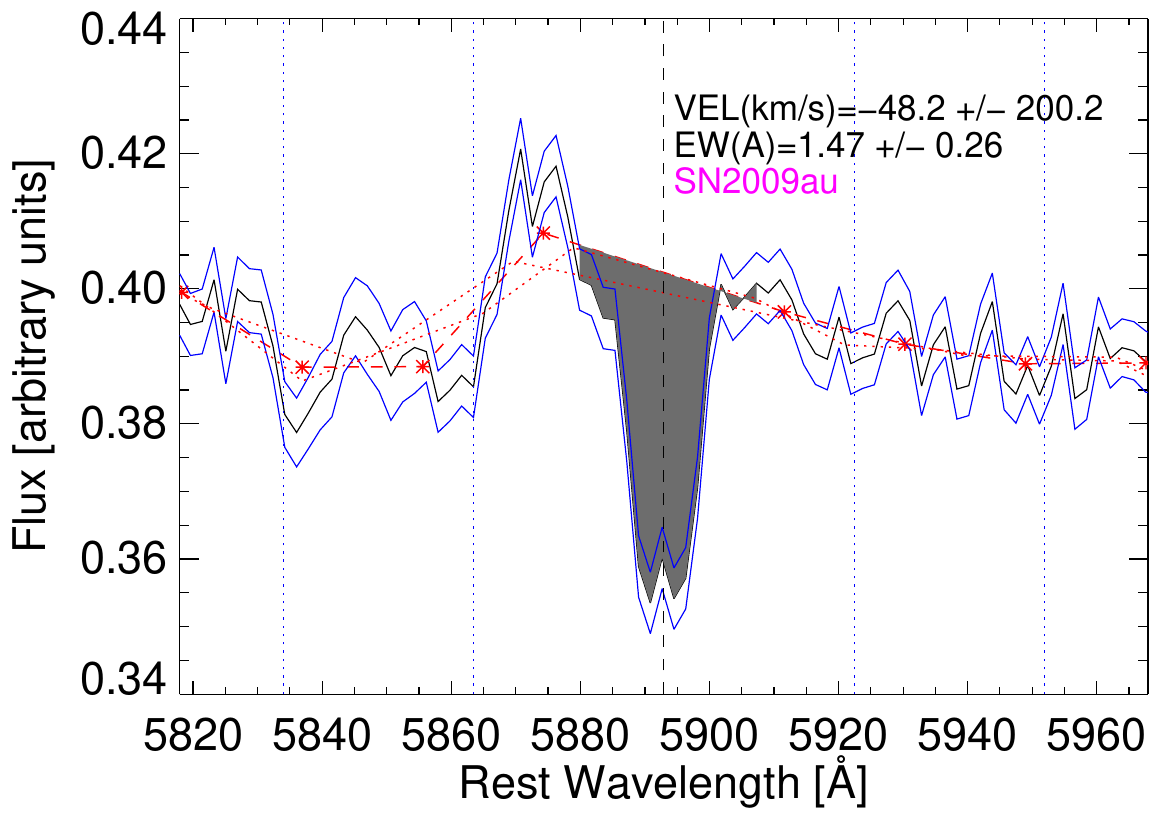}
\caption{SN~2009au (SN~II) spectrum around the \naid lines. The black line is the spectrum, the blue lines are the estimated flux errors, the red stars are the nodes, and the red dashed line is the continuum crossing the nodes. Red dotted lines represent the continua at $\pm25\%$ of the node separation being considered. The grey-shaded area is the integrated line within $\pm$1000 km/s. The black dashed vertical line is the central wavelength of the line. The other vertical blue dotted lines indicate other wavelengths where we measured EW and whose dispersion provides a systematic uncertainty.}
\label{fig:meth}
\end{figure}

To measure the equivalent width of the various line-of-sight narrow lines from the host galaxy and/or the MW, we use a fully automated code inspired by \citet{Forster12}. First, a continuum is found by defining several nodes around the centre wavelength of the narrow line with separations between 50 and 2000 \kms. The nodes are evenly spaced (in velocity space), and the flux at each node is obtained by smoothing locally the spectrum with a cosine kernel. 

Once the continuum is defined, we measure the EW:

\begin{align}\label{eq:ew}
\mathrm{EW} = \sum_i \frac{c_i - f_i}{c_i}\Delta\lambda = \sum_i 1-\frac{f_i}{c_i}\Delta\lambda,
\end{align}
where the flux at each pixel, $f_i$, is compared to the continuum at that pixel $c_i$, and $\Delta\lambda$ is the wavelength separation between pixels in \AA. 

We sum the area under/over the continuum in a fixed velocity range around the central wavelength, similar to \citet{Forster12}. The velocity interval changes for each line that we consider and is given in Table~\ref{tab:lines}. These intervals were estimated by a visual inspection of the spectra with the most pronounced narrow features, e.g. SN~2013fc \citep{Kangas16} or SN~2003cg \citep{Elias-Rosa06}; it is largest for the stronger \naid ($\pm$1000~\kms) which includes two lines. We note that the EW measured in this way sums all contributions in the given range, meaning that, in principle, more than one feature could be present. Additionally, it is important to remember that emission lines are taken into account giving possible negative values of EWs. The net sum should be close to zero when there is only noise.

\begin{figure}
\centering
\includegraphics[width=\columnwidth]{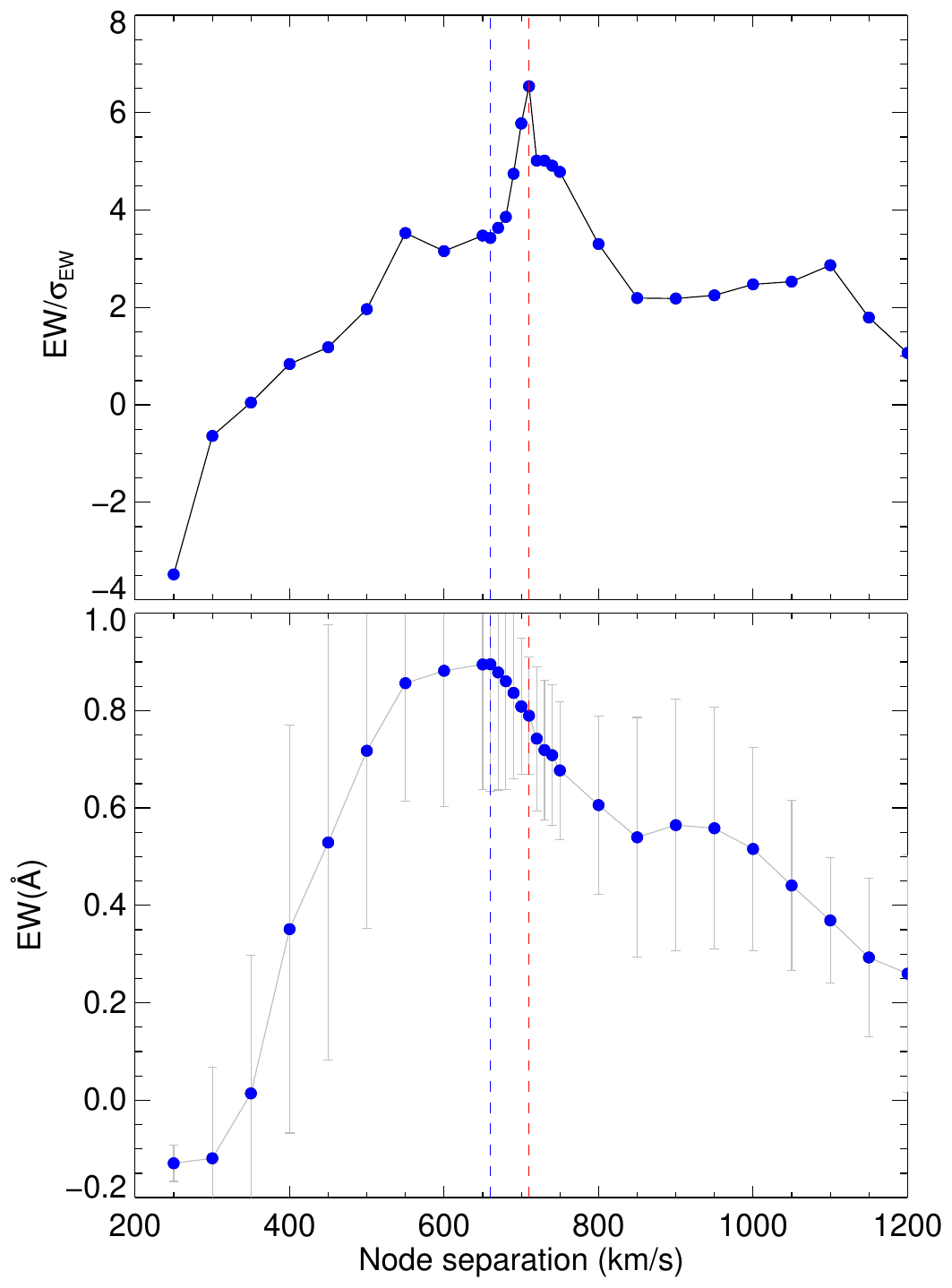}
\caption{Measured equivalent width, EW, over measured uncertainty, $\sigma_{\mathrm{EW}}$, (\textbf{top}) and equivalent width  (\textbf{bottom}) for a spectrum of SN~2014J around the \ki 7665\AA\ (see Figure~\ref{fig:cont}) as a function of the node separation defining the continuum. The maximum EW is shown with a blue dashed vertical line, whereas the maximum |EW|/$\sigma_{\mathrm{EW}}$ is shown with a red line.}
\label{fig:maxsnr}
\end{figure}

The error associated with the EW of a single spectrum has four components: i) the flux error, $\sigma_f$,  which is calculated from the RMS between the flux and the continuum outside the range of the line (but within $\pm$10000 km/s), and an average of this outside error is assumed within the range of the line (see solid blue lines in e.g. Figures~\ref{fig:meth} and \ref{fig:cont}); ii) the error in the continuum, $\sigma_c$, which comes from the dispersion obtained when assuming a difference of $\pm25\%$ in the node separation that determines the continuum (see red dotted lines in Figures~\ref{fig:meth} and \ref{fig:cont}); iii) the error in the integration window for which we re-calculate the EW at $\pm25\%$ of the range given in Table~\ref{tab:lines} for each line; and iv) a systematic error, $\sigma_{\mathrm{sys}}$, from the dispersion obtained when measuring the line at eight different positions where no line is expected (see vertical blue dotted lines in Figures~\ref{fig:meth} and \ref{fig:cont}) and the EW should be zero. The first two sources of error are propagated in Eq.~\ref{eq:ew} to obtain the error on the EW, and together with the latter, they are added in quadrature.

The node separation for the continuum plays a crucial role in this technique (see Figures~\ref{fig:meth} and \ref{fig:cont}), so the optimal separation is found by maximizing the absolute value of the signal-to-noise of the EW (see Figure~\ref{fig:maxsnr}). This is done by calculating the EW for various node separations in bins of 50~\kms\, and through a secondary loop in bins of 10~\kms\, centred at the highest |EW|/$\sigma_{\mathrm{EW}}$ found in the first iteration. The absolute value is used to make sure that we do not necessarily force positive values on the EW. We also require that the profile of the narrow line intersects with the continuum within the velocity window considered.   

\begin{figure}
\centering
\includegraphics[width=\columnwidth]{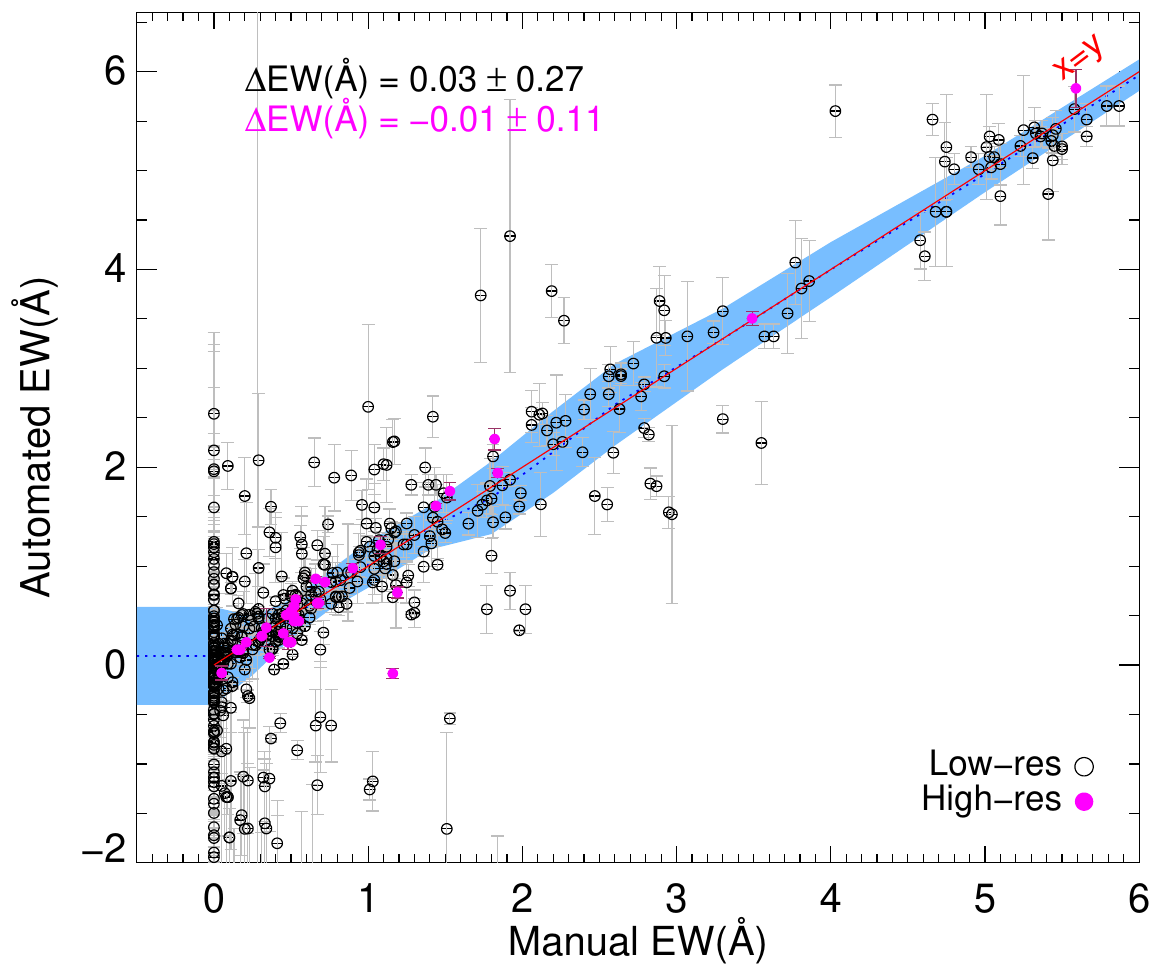}
\caption{Comparison of the EW of \naid doublet measured manually and with our automatic software for a sub-sample of 1145 low-resolution spectra (empty circles) and 32 high-resolution spectra (purple circles). The red line shows a $x=y$ match. The blue shaded area 
represents the median scatter between the two measurements per EW bin. The median and deviation of the full sample are indicated at the top of the plot. The median error from the automatic method is 0.11\AA. Below zero in the manual measurement, we show in blue the median and absolute median deviation, $-0.04\pm0.47$\AA, of the automated measurements for spectra where no line was found visually. }
\label{manual_comp}
\end{figure}

To verify the accuracy of the automated method, we compare the results obtained by measuring the EW of \naid manually using \textsc{iraf} \citep{Tody86_iraf, Tody93_iraf} and with our automatic software. Figure~\ref{manual_comp} shows the comparison for a sub-sample of 1145 low-resolution spectra and 23 high-resolution spectra. The methods give consistent results with a median absolute deviation of 0.01$\pm$0.22 \AA.

\subsection{Velocity measurement} 
\label{sec:vel_meas}

\begin{figure}
\centering
\includegraphics[width=\columnwidth]{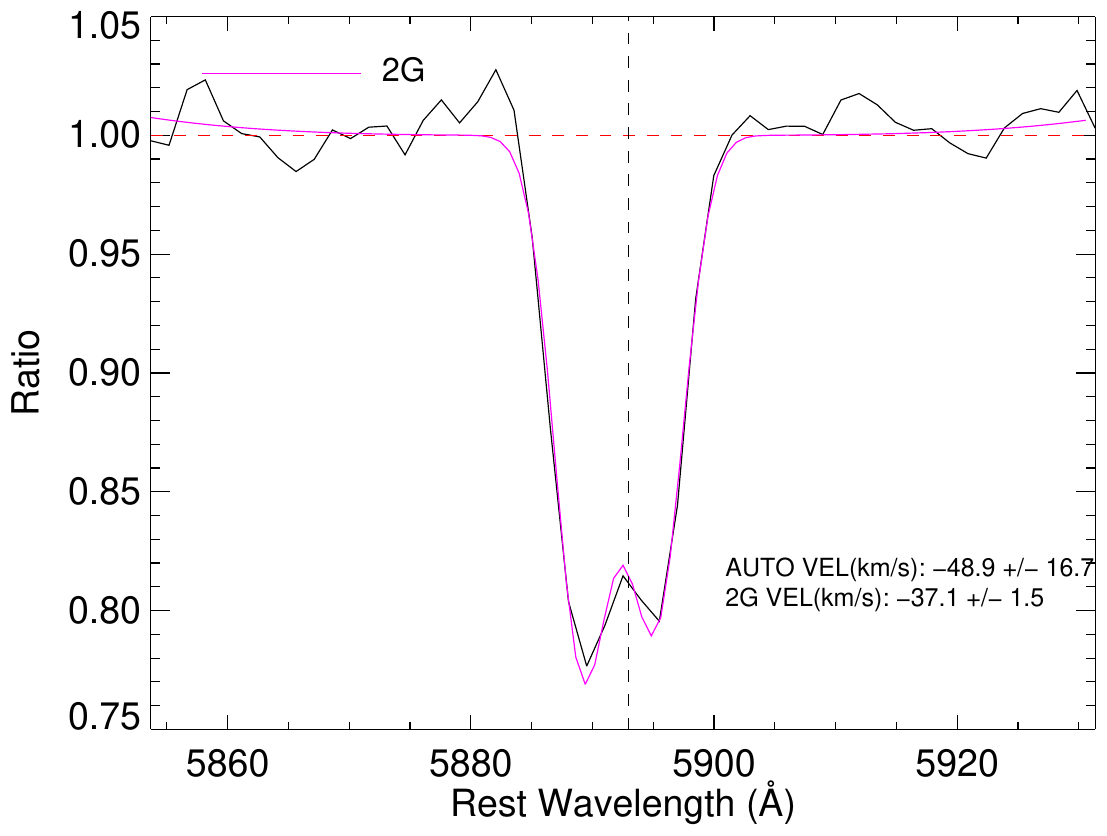}
\caption{Example of a double Gaussian fit (purple) for the host galaxy sodium doublet within a spectrum of SN~2002bo (black). The wavelength separation between both Gaussians is kept fixed. In this case, the fitted velocity is $-37.1\pm1.5$ km/s compared to $-48.9\pm16.7$ km/s (without the uncertainty floor of 35 km/s) obtained with the flux-weighted wavelength average. The vertical dashed line is the mean rest-frame centre of both lines, and the red horizontal dashed line marks the continuum.}
\label{fig:gaussfit}
\end{figure}

The velocity measurement is trickier compared to the EW: for weak or absent lines, it cannot be done, while for stronger lines, there might be multiple components that are only resolvable with high-resolution spectroscopy. Obtaining an unbiased estimate of the velocity similar to the EW is perhaps impossible, but we attempt two approaches. The first is model independent and consists simply of a line-weighted average wavelength of the area under/over the continuum measured above (grey shaded areas in e.g. Figure~\ref{fig:meth}) compared to the rest wavelength of the line:

\begin{align}\label{eq:vel}
\mathrm{v}(\mathrm{km/s}) = \frac{c}{\lambda_{\mathrm{rest}}}\left[\frac{\sum_i (1-f_i/c_i)\lambda_i}{\sum_i (1-f_i/c_i)} - \lambda_{\mathrm{rest}}\right],
\end{align}
where $c$ is the speed of light in vacuum and $\lambda_{\mathrm{rest}}$ the restfame wavelength of the line.

 The first source of uncertainty stems from the wavelength calibration, which may vary substantially across our heterogeneous sample and is impractical to check individually given the lack of, e.g. sky spectra. Instead, we measure the emission lines from the gas phase of the environment, sometimes present in SN spectra, and calculate the average velocity dispersion between those lines. We obtain a characteristic dispersion of 35 km/s (see Appendix~\ref{ap:wavecalib}), which we take as an uncertainty floor from the wavelength calibration. Additional sources of uncertainty come from re-measuring the velocities when i) the flux error is added/subtracted to the original flux, ii) when the continuum is changed by $\pm$25\%, and iii) when the integration window is changed by $\pm$25\%. 

This simple technique works well when the narrow line is relatively symmetric but otherwise suffers from several biases. For example, in the case of the \naid, if one of the two doublet components is stronger, the flux-weighted wavelength will be affected, as seen for SN~2002bo (see Figure~\ref{fig:gaussfit}). The bias may be even stronger in the case of multiple components with different velocities.

\begin{figure}
\centering
\includegraphics[width=\columnwidth]{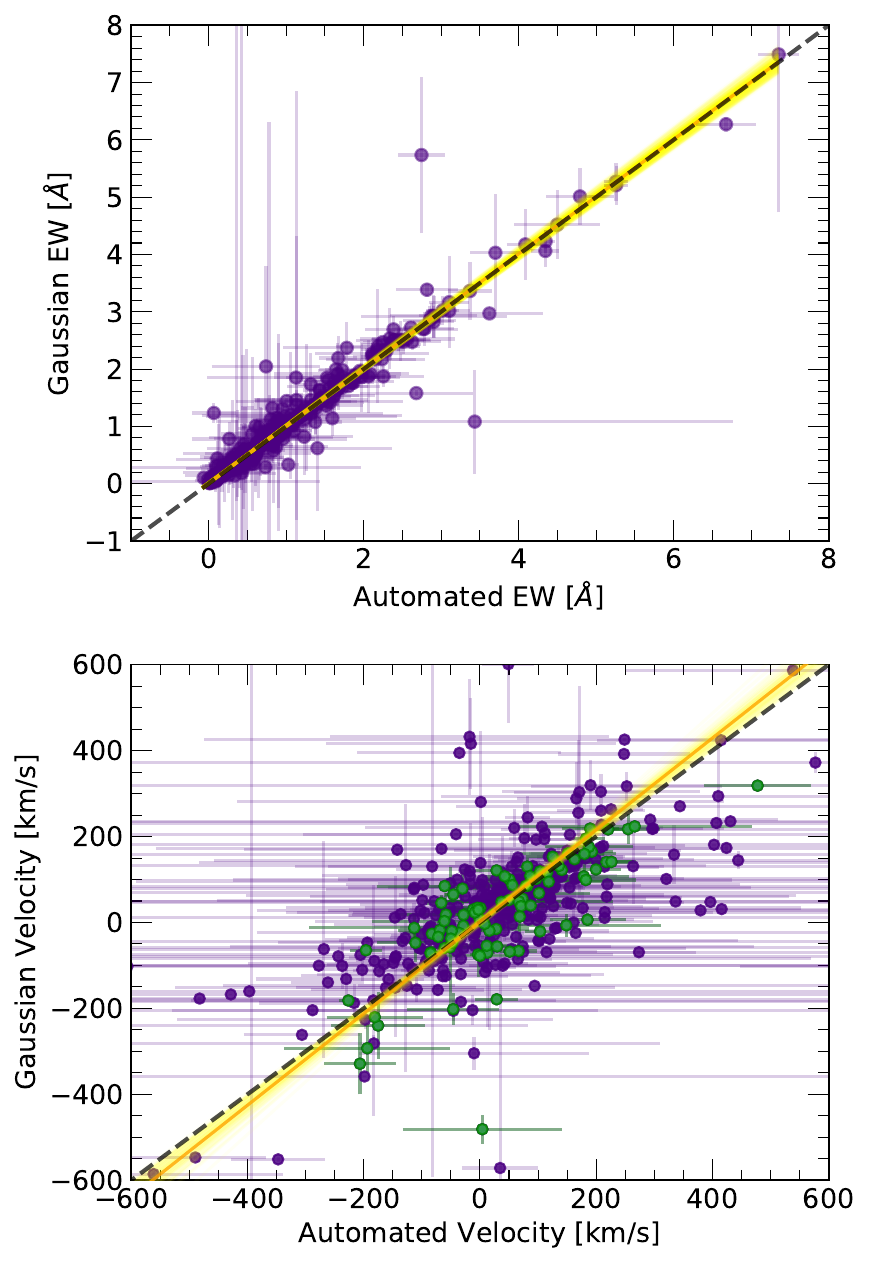}
\caption{Comparison of EWs (\textbf{top}) and velocities (\textbf{bottom}) measured with the automated technique and a double Gaussian fit for \naid with the wavelength separation fixed. Each plot shows the corresponding best linear MCMC fit (\textsc{linmix\_err}; \citealt{Kelly07}) as a solid orange line and the variance with respect to the fit line in yellow. The $x=y$ line is indicated in dashed green lines. In the bottom plot, the green points show that cases for which the EW/$\sigma_{\mathrm{EW}}>1.3$ and EW$>0.5$ have lower uncertainties and agree better within the two methods. The uncertainties here do not include the uncertainty floor of 35 km/s.}
\label{fig:gausscomp}
\end{figure}

Another approach, commonly used in the literature, consists of a parametric profile fitting with, e.g. one or multiple Gaussians. For this, one has to assume the number of components (i.e. number of Gaussians), which one may not be able to distinguish in the case of low-resolution spectra. In Figure~\ref{fig:gaussfit}, we show a fit with two Gaussians with a fixed wavelength separation of 6 \AA\ between the two sodium lines and a common Gaussian width. We directly fit the ratio of the flux-to-continuum using the continuum obtained in the previous section. For this particular case, the line-weighted average estimates a slightly bluer velocity because of the stronger line of the doublet. A comparison of the automated EW and flux-weighted velocity measurements of the sodium doublet with the values obtained from a double Gaussian fit is shown in Figure~\ref{fig:gausscomp}. The match of the EWs is remarkable, while there is a larger dispersion for the velocity -- still consistent within the large uncertainties. This dispersion is mostly driven by cases where the EW is very weak, and the measurement of the velocity becomes very uncertain for both methods. Spectra with stronger lines (see green points) agree better and present lower uncertainties. The two approaches are overall quite consistent, and we use our automated technique as our primary method but compare, when necessary, with the Gaussian approach. 

\subsection{SN average of multiple spectra}
\label{sec:multmeas}

\begin{figure}
\centering
\includegraphics[width=\columnwidth]{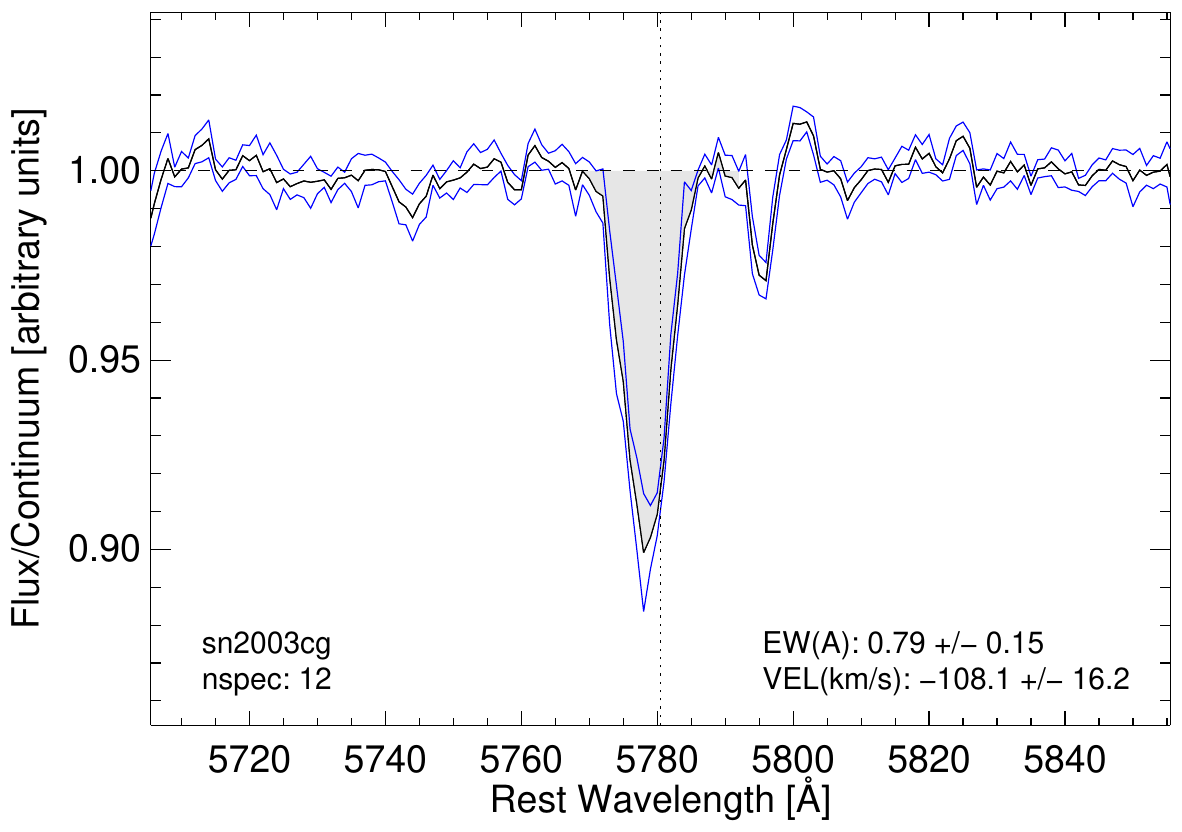}
\caption{Stacked flux-to-continuum ratios of 12 spectra of the SN~Ia SN~2003cg around the DIB-5780\AA\, line (black) with the 1$\sigma$ uncertainty from 100 bootstrap realizations (blue). The EW and velocity values measured with the automated technique (grey shading) are shown in the plot. The vertical dotted line indicates the wavelength of the line.}
\label{fig:stack}
\end{figure}

In the case where multiple spectra of a SN are available, and no evolution is expected or found (see section~\ref{sec:evolution}), one can get a better estimate of the EW and velocity of the intervening line by averaging the individual values found in the previous sections (sections~\ref{sec:ew_meas} and \ref{sec:vel_meas}). Nonetheless, given the heterogeneous set of spectra taken with different instruments and with different underlying SN ejecta throughout the SN evolution, we find that a better approach consists of taking the median of the individual flux-to-continuum ratios to generate a stacked ratio with higher signal-to-noise. This approach is similar to the work done by \citet{Poznanski12} for \naid measured in galaxy spectra. We show an example of 12 stacked flux-to-continuum spectra of SN~2003cg around the DIB-5780 \AA\ line in Figure~\ref{fig:stack}. After the stacking, the EW and velocity are calculated in the same way as for individual spectra. To measure the corresponding error of this stacked value, we do a bootstrap with 100 iterations in which the same number of flux-to-continuum spectra are randomly taken from the original sample (some spectra may be repeated or missing in each iteration), and the EW and velocity are measured each time. The final asymmetric uncertainties are given by 16\% and 84\% of the distribution. The blue lines in Figure~\ref{fig:stack} contain the 1$\sigma$ bootstrap realizations for illustration purposes. As for individual spectra, we also include the systematic errors from the continuum, the integration window and the dispersion of measuring the line at other wavelengths. A comparison of the EW measured from stacked spectra with the median or weighted averages of individual EW measurements is shown in Figure~\ref{fig:stackcomp}.

\section{Spurious temporal evolution of narrow lines and systematic biases}
\label{sec:bias}

\begin{figure}
\centering
\includegraphics[width=\columnwidth]{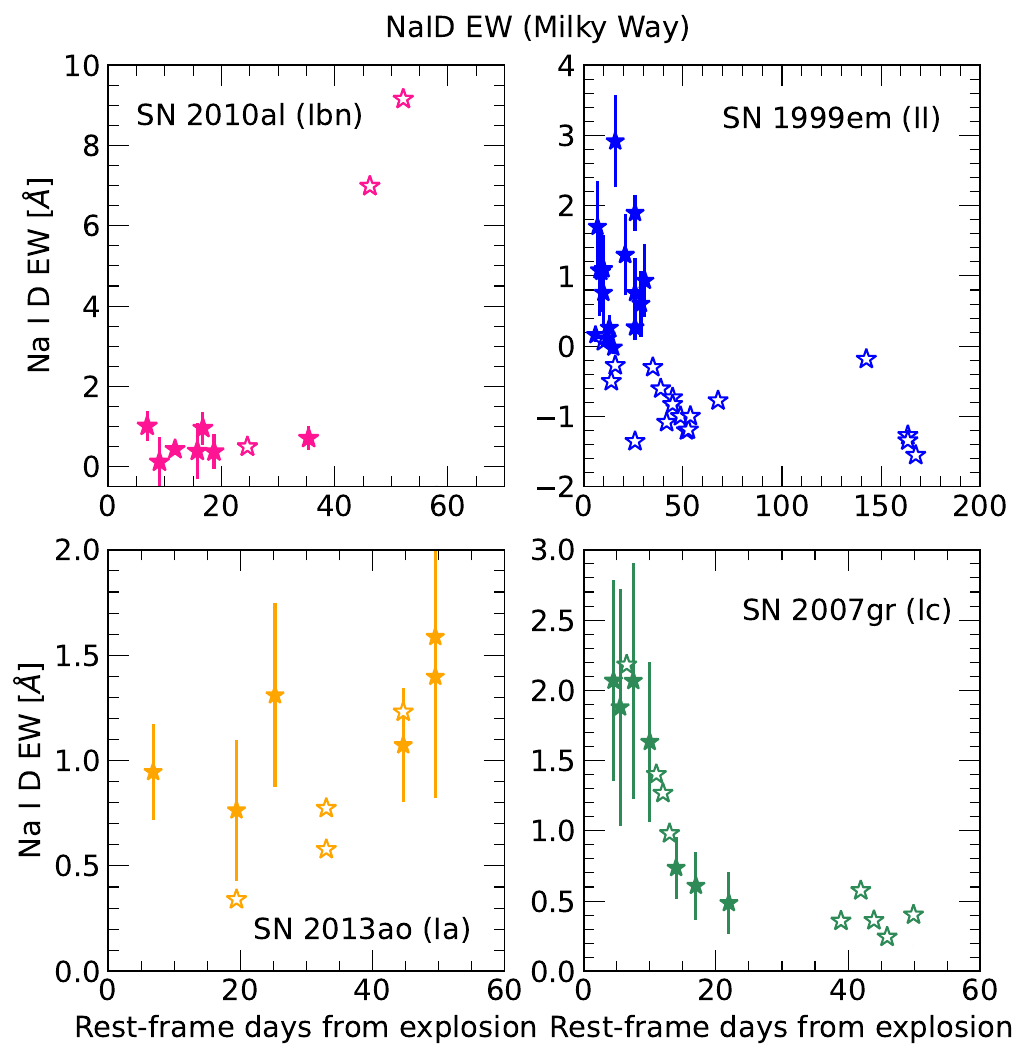}
\caption{\naid\ EW measured at different epochs since explosion for the MW narrow line-of-sight line. Each panel shows the SN name and its type. Open symbols show all available measurements whereas filled symbols are the ones passing the cuts (see section~\ref{sec:bias}).}
\label{fig:EWMW}
\end{figure}

Although the methodology laid out in the previous section correctly measures the EW of narrow lines of spectra of various resolutions, there are some general caveats inherent to the use of low-resolution spectra that we investigate further. Some of these biases become evident when studying the behaviour of the narrow lines over time: we see a conspicuous evolution of the EW for some SNe. In particular, in Figure~\ref{fig:EWMW}, we show the evolution of the \naid\ EW for the MW line for different SN types measured with the automated technique. All objects show variations in their EWs; however, in SN~1999em and SN~2007gr, a constant decrease is striking. Narrow intervening lines from the MW are expected to be generally constant as a function of time since they come from material in our own Galaxy that does not interact with the SN; therefore, variations in their EWs suggest the interference of external factors. A similar, albeit more extreme, behaviour was observed for the narrow lines from the host galaxy (see section~\ref{sec:broadline}). This is worrying since it can lead to false physical interpretations like interaction with circumstellar material.  These results motivate us to explore the possible biases that affect the strength of the narrow line-of-sight lines. They are described in detail in the following subsections.

\begin{table*}
\begin{threeparttable}
\centering
\small
\caption{Number of SNe and spectra after various cuts.}
\label{tab:cuts}
\begin{tabular}{m{3.0cm} cc cc cc cc cc c}
\hline
Cut & \multicolumn{2}{c}{SNe Ia} & \multicolumn{2}{c}{SNe II} & \multicolumn{2}{c}{SE SNe} & \multicolumn{2}{c}{SNe-int$^{\ast}$} & \multicolumn{2}{c}{Total} & Comment\\
& $N_{\rm SN}$ & $N_{\rm spec}$ & $N_{\rm SN}$ & $N_{\rm spec}$ & $N_{\rm SN}$ & $N_{\rm spec}$ & $N_{\rm SN}$ & $N_{\rm spec}$ & $N_{\rm SN}$ & $N_{\rm spec}$ & \\
\hline
None                   & 1017 & 6397 & 327 & 2538 & 328 & 2053 & 116 & 1269 & 1788 & 12257 &  --                        \\
$z>0.004$              & 988 & 5759 & 285 & 1890 & 303 & 1694 & 110 & 1177 & 1686 & 10520 & Section~\ref{sec:z}         \\
S/N$>15$               & 988 & 4171 & 285 & 1515 & 303 & 1099  & 110 & 834 & 1686 & 7619 & Section~\ref{sec:signois}   \\
$m_c<0.002~$\AA$^{-1}$ & 701 & 2466 & 214 & 809  & 233 & 856  & 98 & 439 & 1246 & 4570 & Section~\ref{sec:broadline}$^{\dagger}$ \\
$N_{\rm sp}>1$         & 400 & 2165 & 140 & 735 & 142  & 765 & 71 & 412 & 753 & 4077 & Section~\ref{sec:nspec} \\ 
\hline
\end{tabular}
\begin{tablenotes}
\small
\item $^{\ast}$ These include interacting SNe: SNe~IIn, Ibn and Icn.
\item $^{\dagger}$ Values for \naid. The exact cuts and number of SNe/spectra changes according to line species. 
\end{tablenotes}
\end{threeparttable}

\end{table*}

\subsection{Redshift}\label{sec:z}
In order to ensure that the narrow \naid absorption line from the MW and the host galaxy are separately resolved (i.e., to ensure they are not blended), we set a lower limit cut at 0.004 on the redshift of our sample (independently of the spectral resolution). This eliminates a small number of very nearby SNe (see Table~\ref{tab:cuts}). As will be shown later in section~\ref{sec:broadline}, the redshift also plays an important role in the location of the broad lines of SNe with respect to the narrow intervening lines affecting their measurement.

\subsection{Spectral resolution}
\label{sec:specres}

\begin{figure}
\centering
\includegraphics[width=\columnwidth]{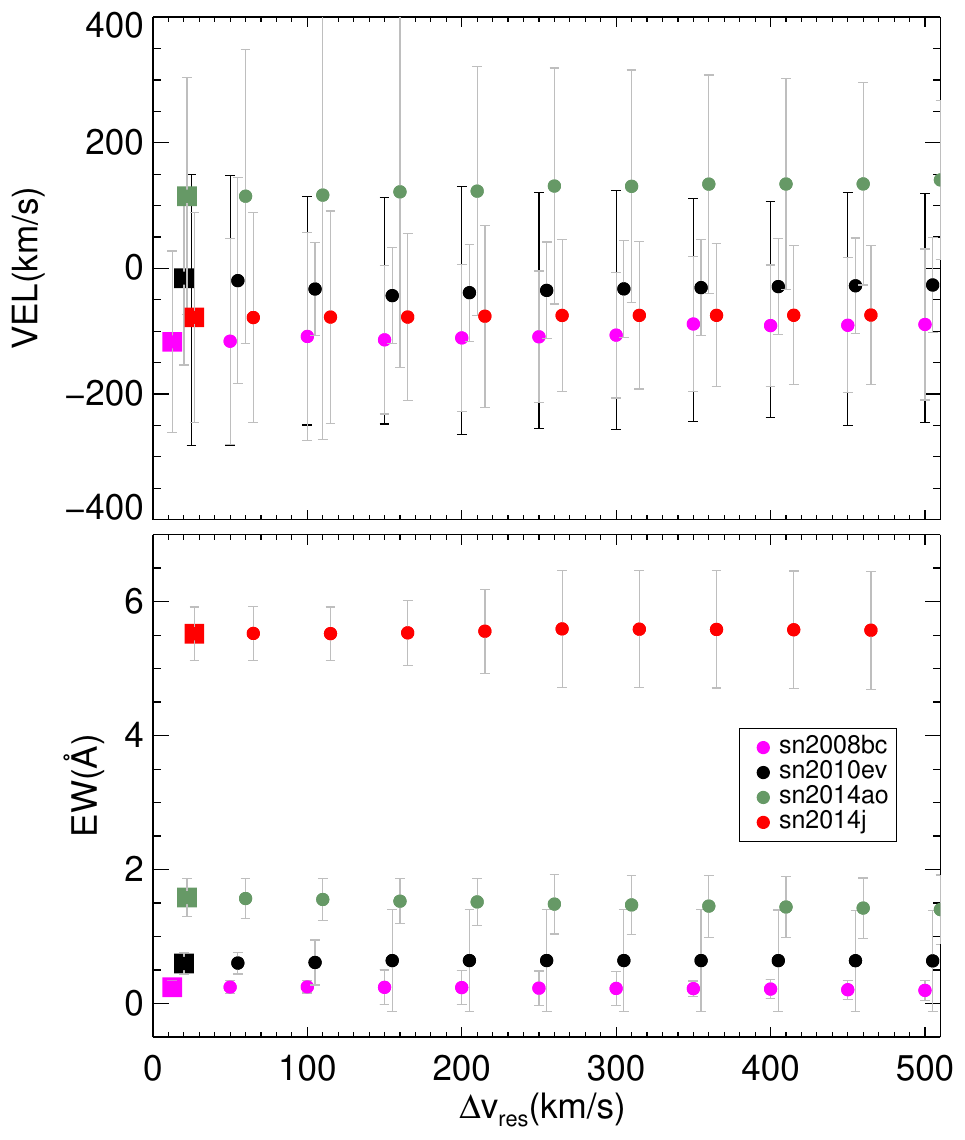}
\caption{Measured velocity \textbf{(top)} and equivalent width \textbf{(bottom)} of \naid as a function of the decreasing simulated resolution, $\Delta \mathrm{v}_{\mathrm{res}}$ for the SN~2010ev spectra shown in Figure~\ref{fig:specres_sim} (black) and also for smoothed high-resolution spectra of SN~2008bc (purple), SN~2014ao (green) and SN~2014J (red). The squares show the original high-resolution values of each SN.}
\label{fig:specres}
\end{figure}

In this section, we explore the impact of varying spectral resolutions in measuring the EW and velocity of the lines from intervening matter. We pick high-resolution spectra ($\Delta v_{\mathrm{res}}<50$~\kms) and smooth them with a cosine kernel (see Section~\ref{sec:measurements}) with increasing velocity windows to simulate lower resolutions (Figure~\ref{fig:specres_sim}). We try velocity windows spanning 50-500~\kms, which conservatively cover our sample's average velocity resolutions (see Figure~\ref{histograms_sigres}). As long as the continuum is relatively flat (Section~\ref{sec:broadline}), we find excellent agreement in the measured EW and velocity of the line-of-sight line for all resolution windows with the original high-resolution values. This is shown in Figure~\ref{fig:specres} for several SNe of different \naid line strengths. We highlight that the uncertainties in the EW become larger, as expected, for worse resolutions. Therefore, no cut on the resolution is applied to our sample. 

\subsection{Signal-to-noise}
\label{sec:signois}

\begin{figure}
\centering
\includegraphics[width=\columnwidth]{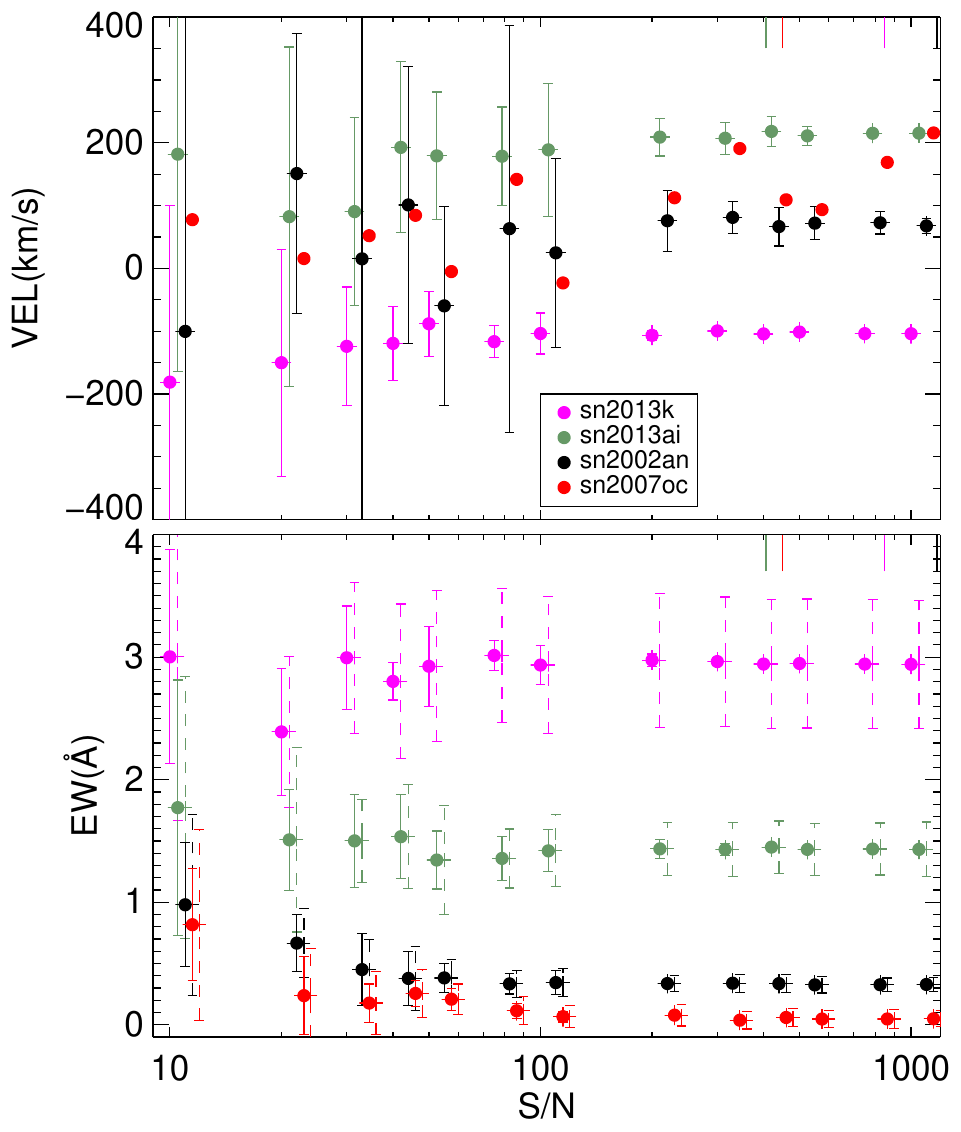}
\caption{Measured equivalent width \textbf{(bottom)} and velocity \textbf{(top)} as a function of the S/N for the perturbed SN~2002an spectra shown in Figure~\ref{fig:signois_sim} (black) and also for perturbed spectra of SN~2013K (purple), SN~2013ai (green) and SN~2007oc (red). The points are the mean of 10 different simulations for a given S/N, and its associated error bars are the standard deviations (SN~2007oc has too large errors on the velocities and are thus not shown). The dashed error bars of the EW are the mean of the errors calculated from the automated technique. Upper right ticks show the S/N of the original unperturbed spectra.}
\label{fig:signois}
\end{figure}

We now evaluate the performance of the automated technique when subject to varying S/N ratios in the spectral range of the narrow line. The S/N, as explained in section~\ref{sec:sample}, is estimated by calculating the RMS of the observed spectrum with respect to a smoothed version. We perform a simulation by taking spectra with high signal-to-noise and randomly perturbing them according to various decreasing S/N ratios (see Figure~\ref{fig:signois_sim}). For each given S/N, we do 10 simulations and take the mean and standard deviation to show the effect in Figure~\ref{fig:signois}. The measured EWs are remarkably consistent with decreasing S/N, and the uncertainties measured by the automated code (dashed error bars) are always larger than the standard deviation (solid error bars) of the simulations. Nevertheless, we do see that for weak lines (e.g. those in the spectra of SN~2002an and SN~2007oc), the average EW starts to increase at very low S/N, albeit always within the errors. Because of this, we add a conservative additional minimal S/N cut of 15 that rejects around 1500 spectra (see Table~\ref{tab:cuts}).

The recovered velocities are also consistent, but we note that, as expected, there is a large dispersion for weak lines as there is not much line flux to obtain the flux-weighted average wavelength. However, the uncertainties on the velocities are always much larger than the standard deviations of the simulated spectra (measured uncertainties are larger than 300~\kms\ and are not shown in the Figure).

\subsection{Broad line interference}
\label{sec:broadline}

\begin{figure}
\centering
\includegraphics[width=\columnwidth]{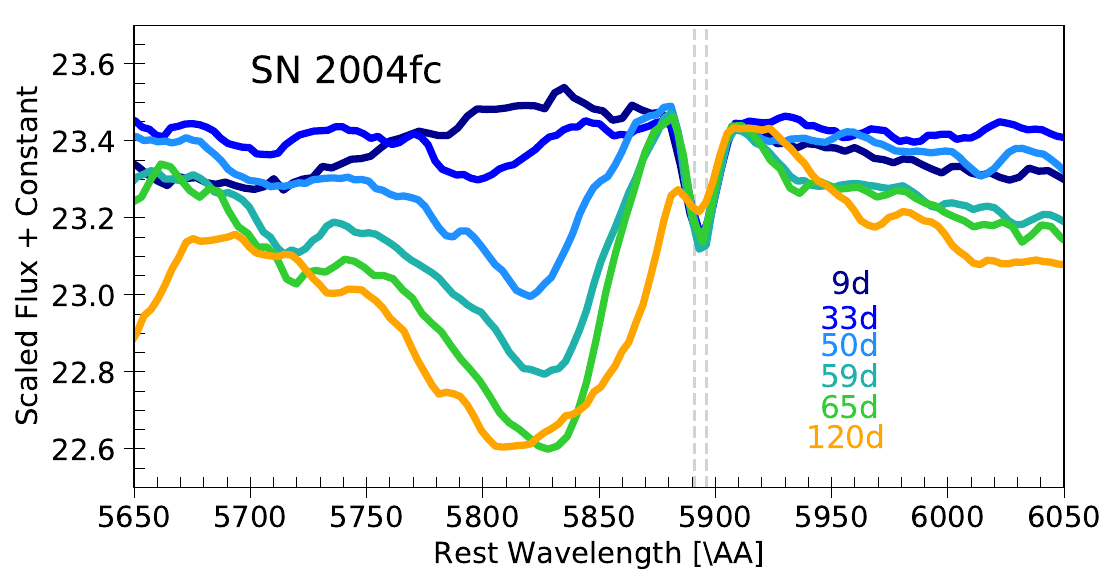}
\includegraphics[width=\columnwidth]{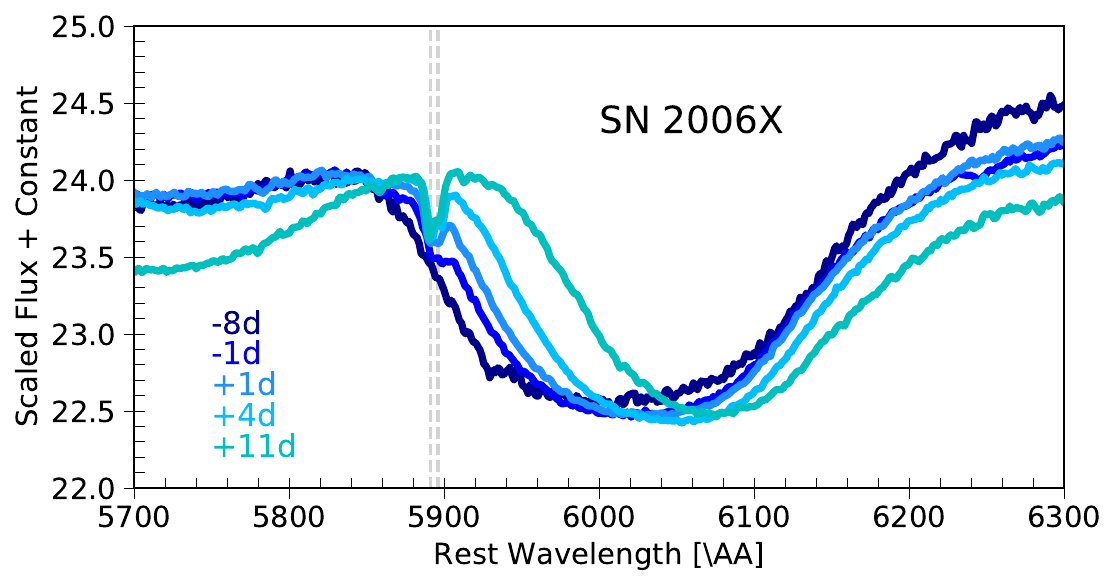}
\caption{Optical spectral evolution of SN~II 2004fc \textbf{(top)} and SN~Ia 2006X \textbf{(bottom)} around the \naid doublet. An apparent evolution of the narrow line is seen for both cases that coincide with the P-Cygni profile evolution of \naid at the left of the narrow line and of Si II at the right of the narrow line, respectively.}
\label{fig:fake_evol}
\end{figure}

When measuring the narrow line-of-sight lines in individual spectra of SNe and investigating their evolution, it becomes evident that several objects have a seemingly temporal change in the strength of \naid. Particularly for SNe~II, there is often a strong apparent decrease in the EW as a function of time, but variations can also be seen for SNe~Ia (see, e.g. Figure~\ref{fig:fake_evol}). Upon closer examination of the SN spectra, these changes appear to be strongly linked to the appearance and evolution of the emission and absorption components in the P-Cygni profile of the broad \naid (for SNe~II) and \sii 6355 \AA\ (for SNe~Ia). If the variations are connected to the development of broad lines (see \citealt{Gutierrez17a} for the appearance and evolution of P-Cygni lines in SNe~II), they do not constitute real intrinsic variations. In that case, one will infer a wrong evolution and, more generally, EW measurements that depend on the underlying profile and are thus not a reliable reflection of the column density of the intervening material.

To investigate such bias in more detail, we create a simple simulation in which we input fake narrow lines into various differing P-Cygni profiles and calculate the recovered EW. For these simulations, we take both observed optical spectra of seven SNe~II, three SNe~Ia, and five SESNe, together with two SN radiative transfer models: the delayed-detonation model of SNe~Ia \citep{Blondin15}, and the m15mlt3 for SNe~II \citep{Dessart11, Dessart14}. Most of the selected observed SNe do not show evident narrow \naid lines; however, when the lines are clearly detected, they are removed from the spectra by interpolating the continuum to the line wavelengths so that we can then add simulated lines of our choice. 

Once the spectra are clean from the \naid lines, we smooth them using a moving average box by convolution with a cosine kernel. Although most spectra have a good S/N, the smoothing process provides easier identification of the continuum and the narrow lines. 
Following the smoothing, the spectra are normalised at the average position of the narrow lines ($\lambda5893$). In parallel, we generate narrow \naid absorption features with similar characteristics to those observed in SN spectra. These features are created for two cases: i) for low-resolution spectra that assume the \naid doublet is blended and just one single feature is detected, and ii) for high-resolution spectra that produce the \naid doublet (two resolved lines). For the two cases, we test different strengths of the narrow line.

\begin{figure}
\centering
\includegraphics[width=\columnwidth]{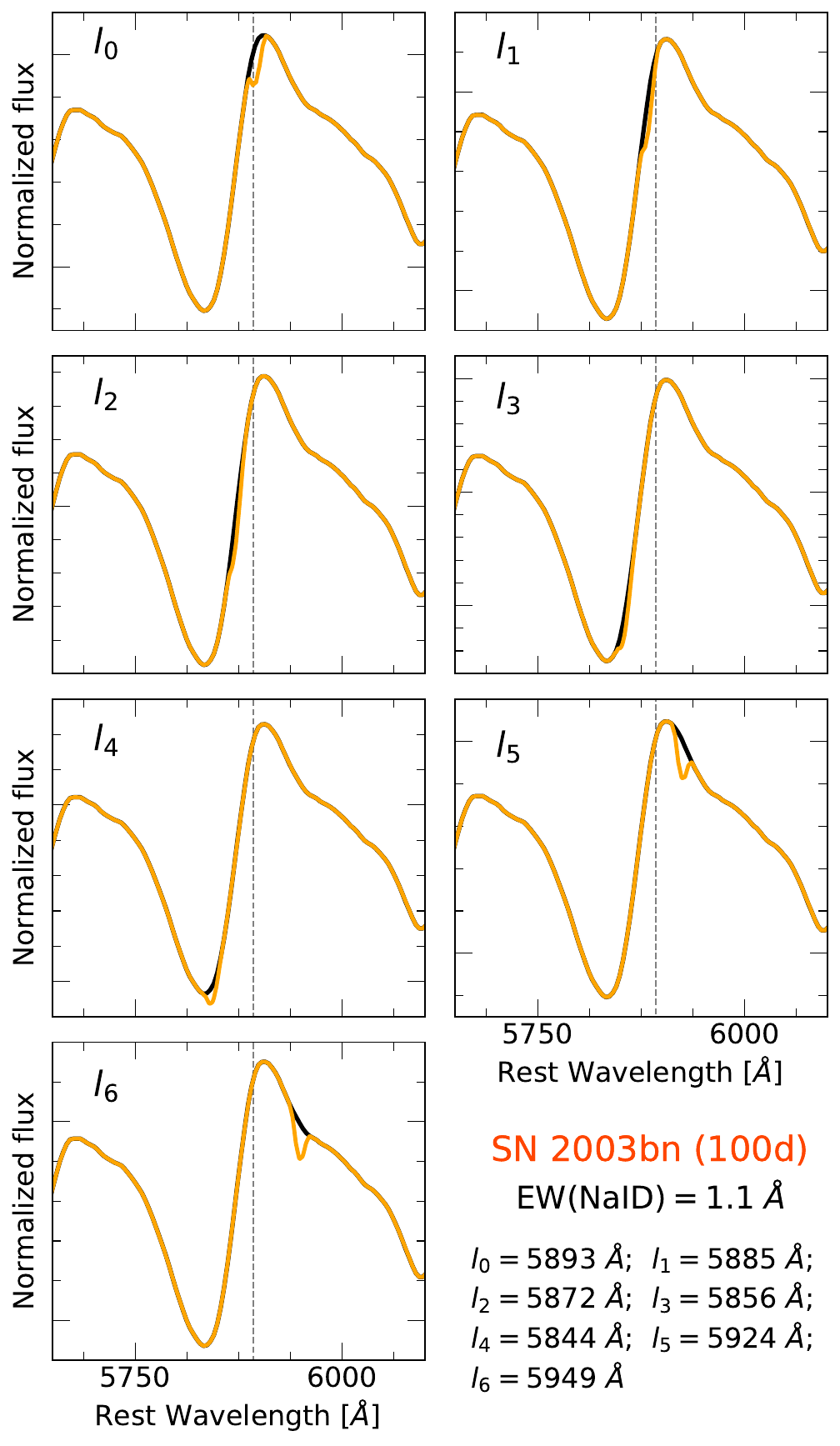}
\caption{Narrow \naid lines artificially placed at various locations for SN~2003bn. Here, we added the narrow \naid lines with low resolution (single line). The location of the narrow line is presented in the different panels.}
\label{mosaic}
\end{figure}

To see the effect of the narrow lines in the SN spectra, we add the narrow \naid feature at different locations offset from the central wavelength in all SN types to simulate different ejecta velocities and redshifts. Figure~\ref{mosaic} shows the type II SN~2003bn spectra at 100 days from explosion centred on the broad \naid P-Cygni profile with the added narrow \naid feature at different locations. Here, the narrow \naid features were added for low (single line) resolution spectra. As one can see, the strength of the narrow \naid lines at low-resolution visually changes depending on the location in the spectrum; for example, at $l_2$ and $l_3$, the line looks weaker. This same procedure of adding the narrow lines at different positions was also done for other observed SNe and for the aforementioned SN~Ia and SN~II models.

\begin{figure*}
\centering
\includegraphics[width=\textwidth]{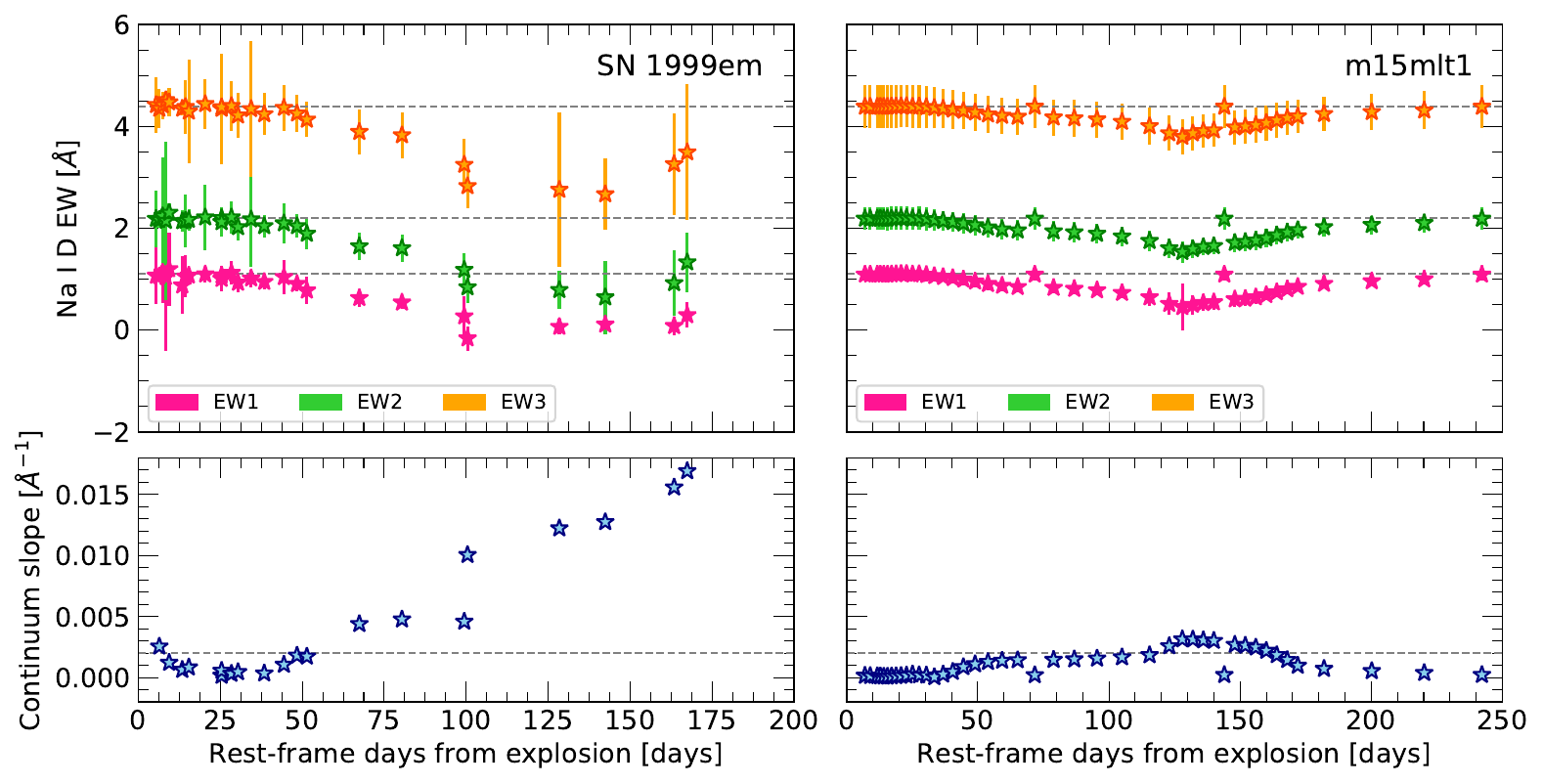}
\caption{EW measurements for simulated intervening \naid lines located at rest-frame ($z_0=5893$ \AA) on top of real sodium P-Cygni profiles from observed spectra of SN~1999em (left) and on top of simulated spectra of \citet{Dessart11} (right). Horizontal lines show the EW expected from the simulated narrow \naid line: $EW1=1.1$ \AA, $EW2=2.2$ \AA\ and $EW3=4.4$ \AA. The measured continuum slope at around 5893 \AA\ is presented in the bottom panel. The horizontal line shows the cut applied in this study: a maximum continuum slope value of 0.002\,\AA$^{-1}$. }
\label{fig:simEW-SNII}
\end{figure*}

When we then calculate the EW of these simulated narrow lines, we find significant changes and a non-intrinsic evolution, as can be seen for observed and modelled spectra of SNe~II in Figure~\ref{fig:simEW-SNII}. It is evident that as the P-Cygni profile of \naid develops with time during the SN photospheric evolution, the narrow line measurement is deeply affected: the EW we measure is lower and drops as a function of the steepness of the underlying continuum. Not only does this affect the EW measured for the narrow line, but it may also affect the photospheric or nebular line from the SN, e.g. producing an apparent redshift in the broad \naid\ emission line at late times in CC-SNe ($>100$ days from explosion; see Figure~\ref{fig:sneII}). This effect depends on the SN cosmological redshift and ejecta expansion velocity, which determine the location of the narrow line with respect to the P-Cygni profile. At different locations, the inferred behaviour changes accordingly. However, if we do the same simulations for high-resolution narrow lines, we do perfectly recover the EW, no matter the underlying P-Cygni profile.

\begin{figure*}
\centering
\includegraphics[width=\textwidth]{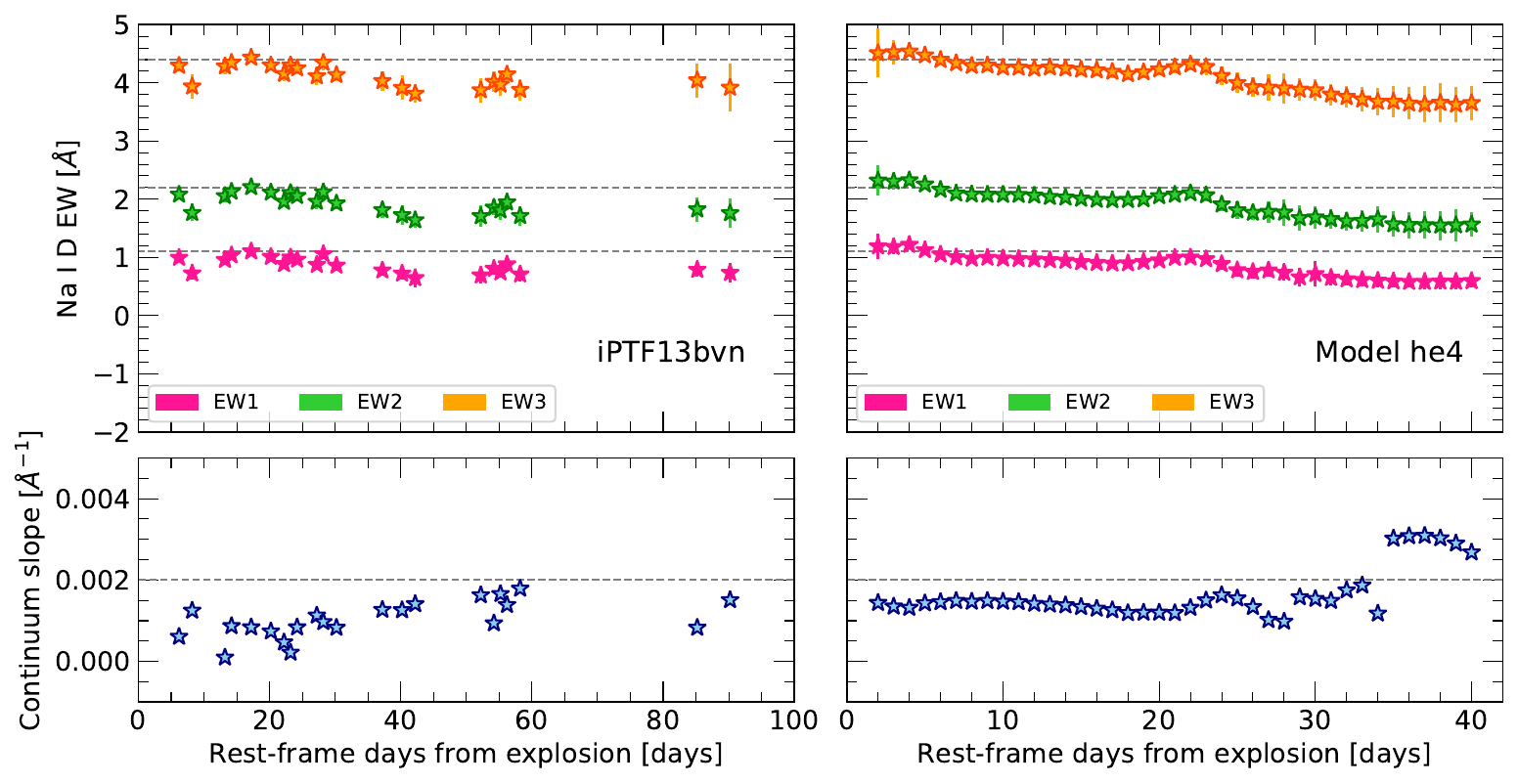}
\caption{Same as Figure~\ref{fig:simEW-SNII}, but for SESNe. Simulated narrow \naid lines on top of observed spectra of iPTF13bvn (left) and on top of modeled spectra of \citet{Dessart20} (right).}
\label{fig:simEW-SESNe}
\end{figure*}

Similarly, we show in Figure~\ref{fig:simEW-SESNe} the results for a simulation of SESN models and observed spectra. We also find that the \naid P-Cygni profile of the SN ejecta may swallow portions of the narrow line at later epochs in low-resolution spectra and consequently induce smaller measurements of the EW. Regarding SNe-int, there are no available models in the literature, and simulating narrow lines on top of a few objects will misrepresent a very heterogeneous class. Nonetheless, we expect --and visually confirm-- the effect to be smaller, since strongly interacting SNe generally do not show prominent P-Cygni profiles.

\begin{figure*}
\centering
\includegraphics[width=\textwidth]{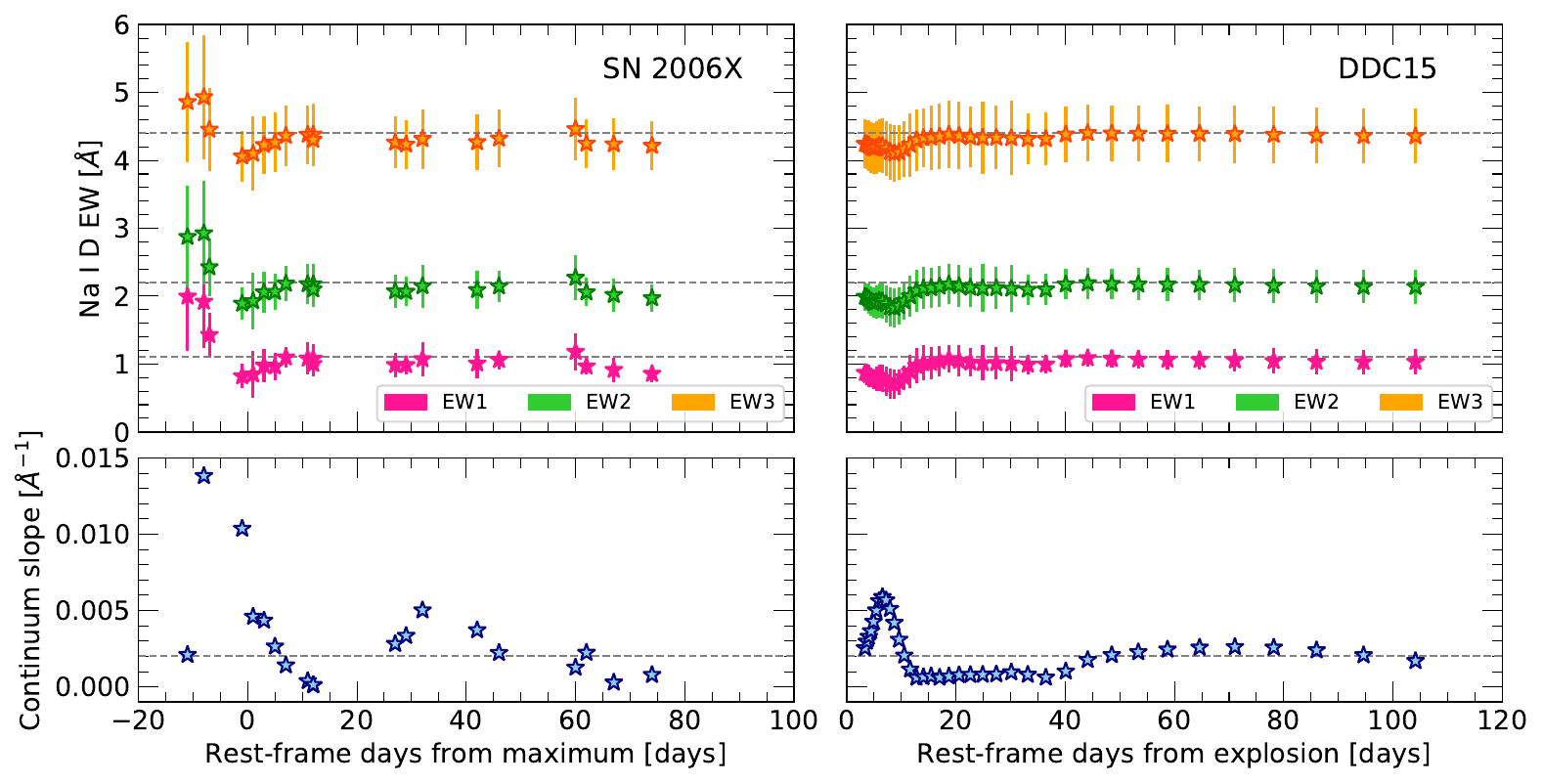}
\caption{Same as Figure~\ref{fig:simEW-SNII}, but for SNe~Ia. Simulated narrow \naid lines on top of observed spectra of SN~2006X (left) and on top of modeled spectra of \citet{Blondin15} (right).}
\label{fig:simEW-SNIa}
\end{figure*}

Although less pronounced than SNe~II, SNe~Ia can also exhibit an apparent evolution from contamination by the broad ejecta features. This is seen for observed and simulated spectra (see Figure~\ref{fig:simEW-SNIa}). In this case, the P-Cygni profile of \sii 6355 \AA\, at early times, has very high expansion velocities, and part of it may reach the rest wavelength of the narrow \naids; later as the ejecta slows down, the narrow line is no longer engulfed by the absorption of the P-Cygni profile, and the EW is well recovered. The exact behaviour naturally depends strongly on the expansion velocity that moves different parts of the broad line into the narrow \naids, and the effect will be stronger for higher-velocity objects (bottom panel of Figure~\ref{fig:fake_evol}). Again, this is not seen for high-resolution spectra. 

It is thus no surprise that using low-dispersion spectra, \citet{Wang19} find the evolution of \naid lines in high-velocity SNe~Ia when not properly considering this effect. Their evolution is strikingly similar to our simulated case shown in Figure~\ref{fig:simEW-SNIa}.  Therefore, variations in the strength of \naid\ narrow lines in SNe~Ia at early times are not always indications of circumstellar interaction. In most cases, when low-resolution spectra are used, variations are produced by the interference of SN broad lines; in this case, by the \ion{Si}{ii} line.

\begin{figure}
\centering
\includegraphics[width=\columnwidth]{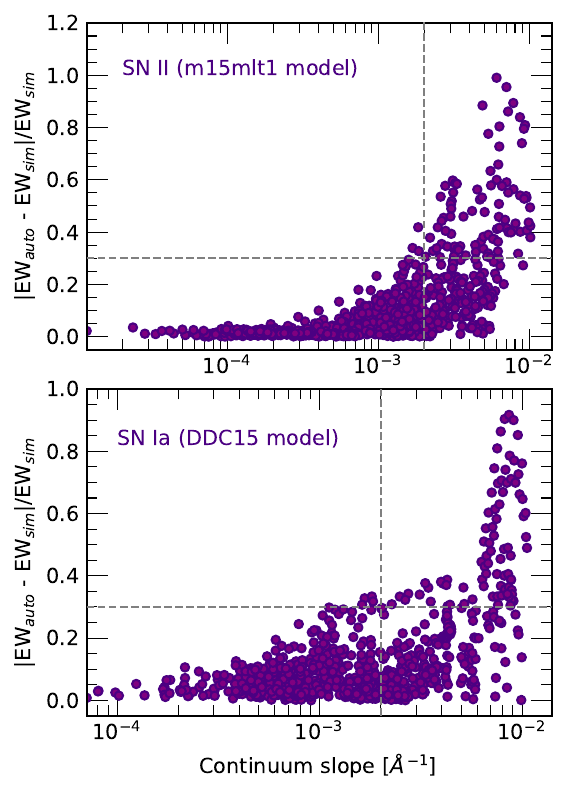}
\caption{Fractional residual of measured EW to simulated EW (|EW$_{\mathrm{auto}}$-EW$_{\mathrm{sim}}|/$EW$_{\mathrm{sim}}$) as a function of the continuum slope around the \naid line for simulated spectra of SNe~II (top) and SNe~Ia (bottom). EW measurements include simulated line-of-sight \naid lines at 6 different positions around the central wavelength and for three different simulated narrow EW strengths. The vertical dashed line is our adopted cut to eliminate spectra with too large a continuum slope, i.e., a big underlying broad line profile, eliminating cases with more than 30\% residuals (horizontal dashed lines). }
\label{fig:dcont-res}
\end{figure}

Having seen the clear important bias of the interference of the broad line in the measurement of the narrow lines, is there a way to identify these dubious cases? Since the bias stems basically from a steep continuum from the P-Cygni profile, we can measure for a given spectrum the average absolute slope of the inferred continuum, $m_c$, from the automatic method (see Section~\ref{sec:ew_meas}) by doing:

\begin{equation}
    m_c = \frac{1}{<f_c>}\frac{1}{N}\sum_i^N \frac{|f_c^{i+1} - f_c^i|}{\lambda_c^{i+1}-\lambda_c^i},
\end{equation}
where the sum is over all N nodes of the continuum within $\pm$50\AA\, of the line, and $<f_c>$ is a normalization average of the continuum flux in the same range. The normalized slope continuum $m_c$[\AA$^{-1}$] is a measure of the steepness of the line in the immediate vicinity of the narrow line. Such slopes clearly reflect the presence of underlying broad profiles, as can be seen in Figures~\ref{fig:simEW-SNII} and ~\ref{fig:simEW-SNIa}. The discrepancy with the real simulated EW increases for larger continuum slopes, as can also be seen in Figure~\ref{fig:dcont-res}. We find that a cut of continuum slopes of $m_c<0.002$ \AA$^{-1}$ ensures deviations of less than 30\% in the EW measurement for all simulated cases. We strongly encourage such a cut in $m_c$ for the EW measurement of narrow lines in mid and low-resolution spectra. 

\subsection{Slit size and orientation}
\label{sec:muse}

\begin{figure}
\centering
\includegraphics[width=\columnwidth]{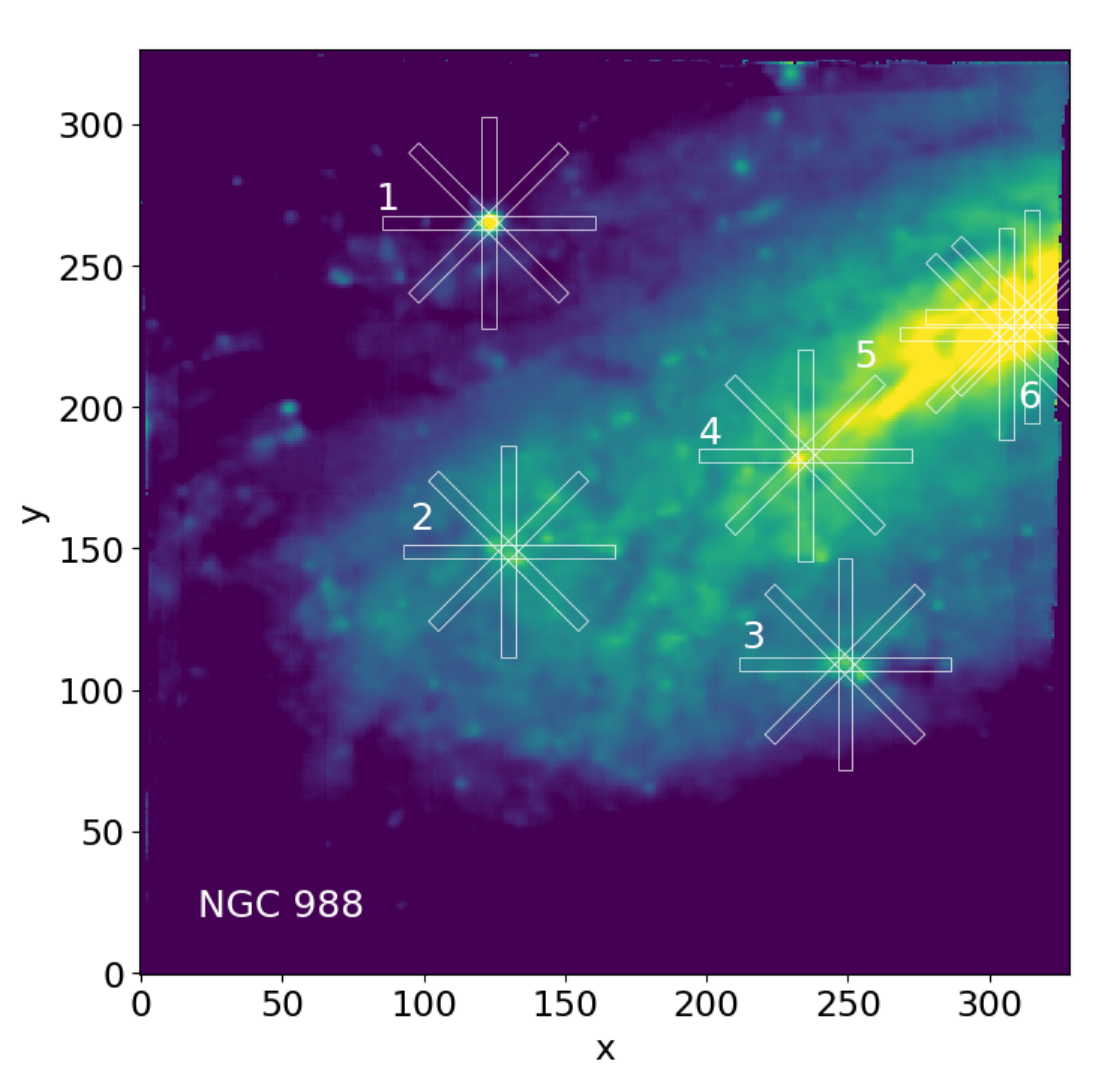}
\caption{MUSE image of the galaxy NGC~988. This synthetic image is created by collapsing the MUSE data cube across the full wavelength range of the observation. Orientation is North-up, East-left. The white rectangles mark four slit angles (0, 45, 90 and 135 degrees) at six different positions, labelled with numbers from 1 to 6. The width of the slit is 1\arcsec~ in this case, but the tests are also performed for a 1.5\arcsec~ slit.}
\label{fig:MUSE}
\end{figure}

The stability of the strength of narrow lines could be affected by different external instrumental factors (besides the resolution and S/N), such as the size and orientation of the slit used for spectroscopy or the amount of host-galaxy light entering into the slit. These factors could introduce changes in the EWs that could be easily confused with intrinsic variations, given the diversity of instruments and observing conditions. To estimate the effect of some of these potential biases in the strength of the narrow lines, we measure the EW of the \naid lines in different environments and with different simulated slits. For this aim, we use data obtained with MUSE at the Very Large Telescope (VLT) as a part of the All-weather MUse Supernova Integral field Nearby Galaxies (AMUSING, \citealt{Galbany16}) collaboration.  
 
\begin{figure}
\centering
\includegraphics[width=0.49\textwidth]{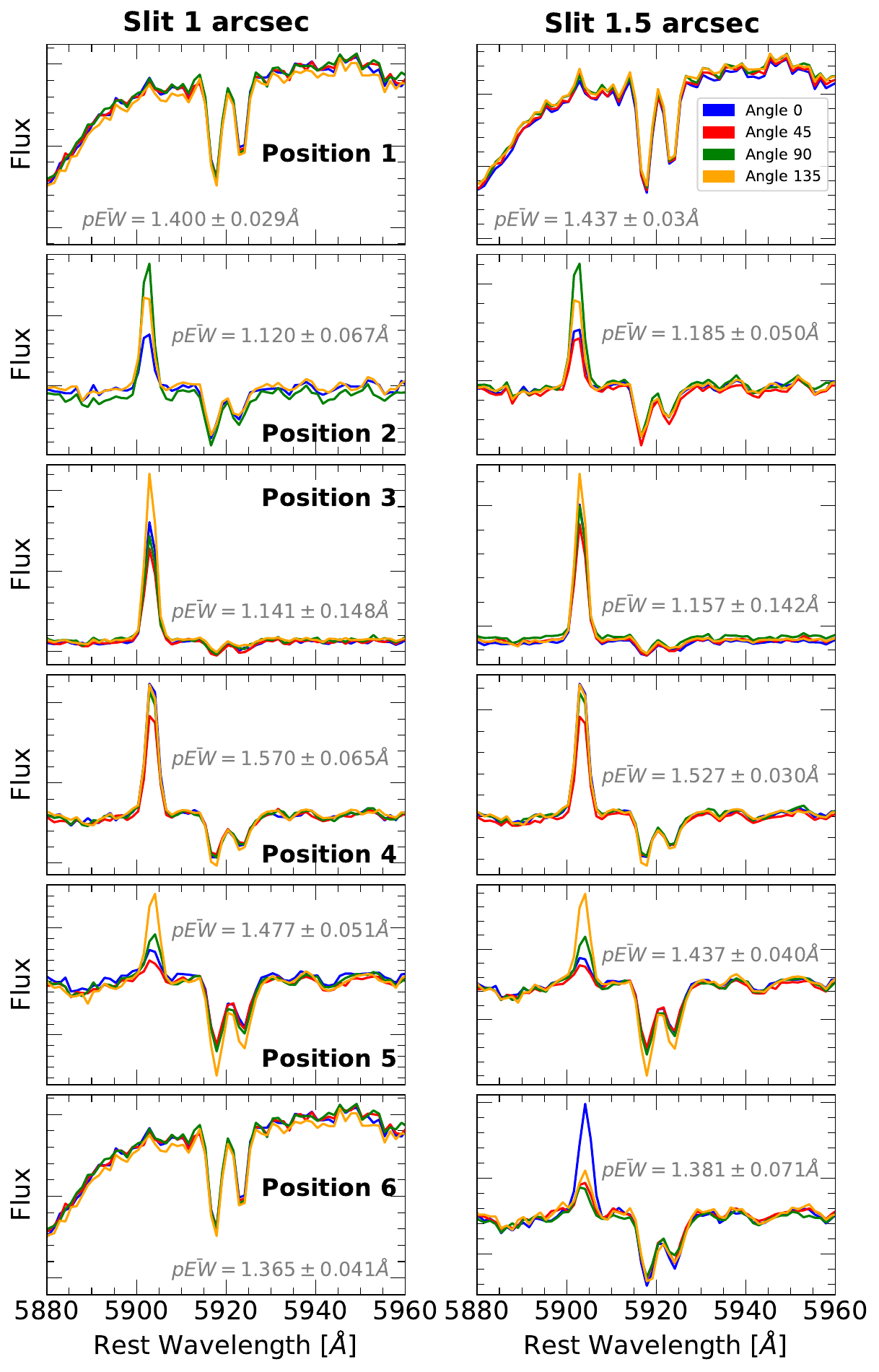}
\caption{Spectra around the \naid line-of-sight line for six slit positions of the MUSE cube shown in Figure~\ref{fig:MUSE} and with four orientations with respect to the horizontal W-E orientation shown in colours: blue ($0^{\circ}$), red ($45^{\circ}$), green ($90^{\circ}$) and yellow ($135^{\circ}$). The left panels are for the 1\arcsec~ and the right panels for the 1.5\arcsec~ slits. The measured EWs and velocities are always consistent within the uncertainties for the different slit orientations and apertures. We note that the emission line to the left of \naid absorption lines is the nebular \ion{He}{I}\, at restframe 5875 \AA.}
\label{fig:Napos}
\end{figure}

To evaluate the effect of the amount of host-galaxy light entering the slit, the impact of the slit orientation and its size, we define six different positions, with four distinct slit angles for each location (0, 45, 90 and 135 degrees with respect to the horizontal West-East axis) and two widths for the slit, one of 1\arcsec~ and another one of 1.5\arcsec. The length of the slit is taken to be 15\arcsec. Figure~\ref{fig:MUSE} shows an example of the procedure for NGC~988, the host galaxy of the type II SN~2017gmr. As one can see, the chosen positions are distributed along the galaxy, from outer regions to the centre. We note that position~1 actually includes the SN~2017gmr in the slit, making this a particularly good study case. We then integrate all spectra within a rectangular aperture of the slit dimension. The resulting profiles centred on the narrow \naid lines are presented in Figure~\ref{fig:Napos} for all different slit orientations and positions. We show the profiles obtained using the slit of 1\arcsec~ width on the left panels and the slit of 1.5\arcsec~ on the right panels. The average and standard deviations of the \naid EWs for every position, displayed in each panel, show small differences (the largest deviation for the slit size is $\sim3\%$ and for orientation is $\sim13\%$ at position~3). Although positions 2-6 do not enclose a supernova, we expect the effect of just integrated host light to be an upper limit on the EW since the light contribution of the SN will dominate. In fact, position~1 with the SN does present only minor changes of less than 3\% with slit size and orientation.
Similar results were also found using several other galaxies targeted by AMUSING, particularly the largest error associated with these components is about 0.20 \AA, which is generally always lower than the uncertainty estimate of our automated technique and thus not a major concern henceforth.

\subsection{Minimum number of spectra}
\label{sec:nspec}

\begin{figure}
\centering
\includegraphics[width=\columnwidth]{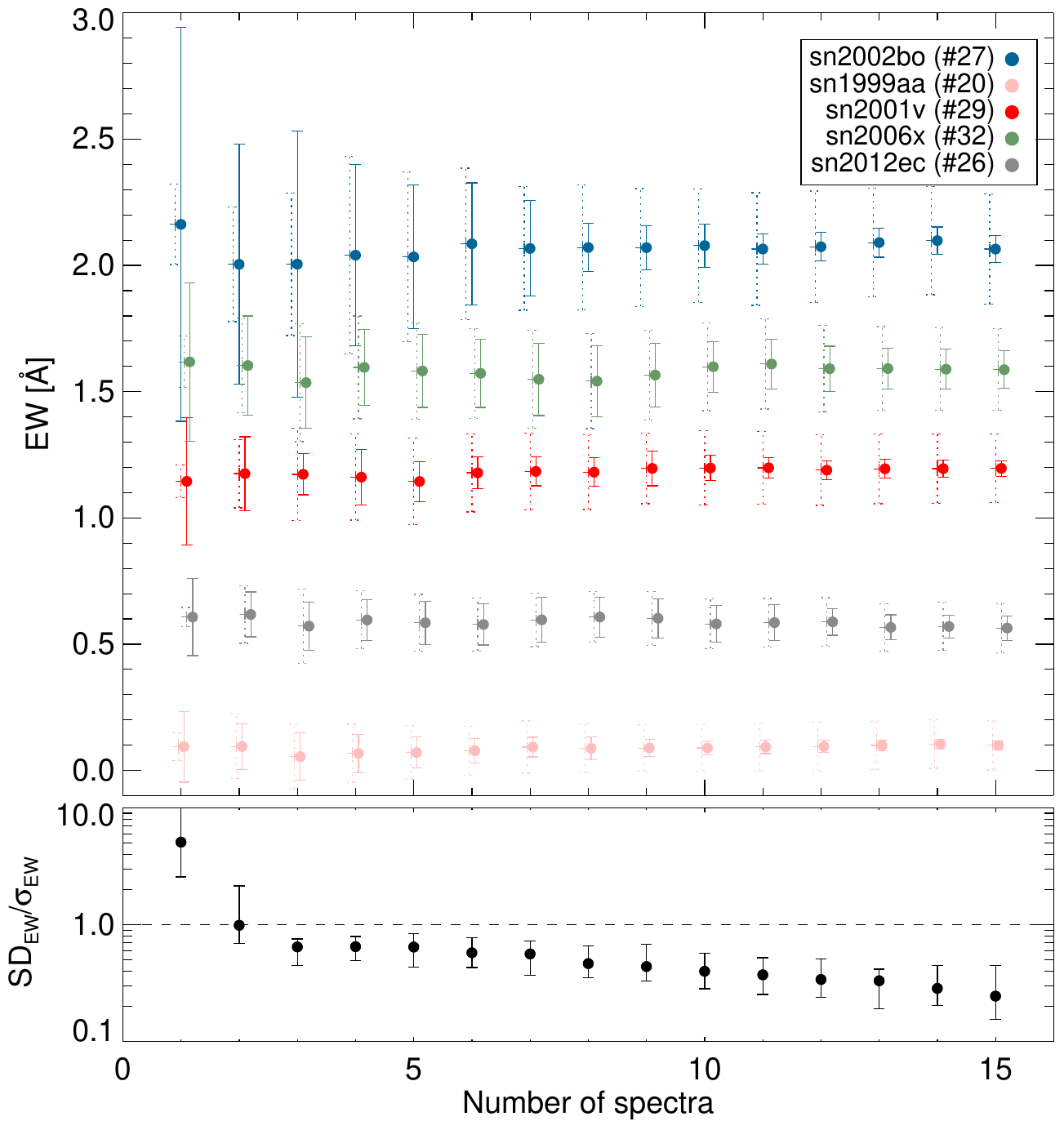}
\caption{\textbf{Top}: Median and standard deviation of equivalent width measurements from multiple spectra randomly taken from the numerous sets of spectra of five SNe: SN~1999aa, SN~2001V, SN~2002bo, SN~2006X and SN~2012ec, as a function of the number of spectra used for the measurement. The solid error bars represent the standard deviation, while the dashed error bars are the median error calculated by the automated technique. \textbf{Bottom}: Average ratio of the standard deviation to the calculated uncertainty from the automated technique for nearly 30 SNe with more than ten spectra. The error bars show the dispersion.}
\label{fig:minspec}
\end{figure}

In this section, we evaluate how the automated measurement of the EW changes with the number of spectra used in the stacking procedure of section~\ref{sec:multmeas}. We take non-evolving SNe with more than 20 spectra, passing the previous cuts (in particular, the continuum slope). We then explore the effect of using 1 to 15 spectra for calculating the EW. For each number of spectra used, we draw 50 random combinations of spectra and calculate the median and standard deviation, as shown in Figure~\ref{fig:minspec}. Besides the dispersion of all draws, we also show the typical median uncertainty obtained from our method with dashed error bars. The bottom plot shows that the uncertainties are conservative when using more than 2-3 spectra. The standard deviation indicates a high dispersion in the measured values for one and sometimes two spectra, even beyond the measured uncertainties. The bottom panel of Figure~\ref{fig:minspec} shows that this underestimation can reach a factor of 2-10 when using only one spectrum. We thus prefer to discard SNe with only one spectrum because of the large dispersion of $1-2$ \AA\ of some cases. 

Our final set of spectra and SNe is considerably reduced after the four cuts (see Table~\ref{tab:cuts}): redshift, signal-to-noise in the spectra, continuum slope, signal-to-noise in the EW and minimum number of spectra. However, we believe these cuts provide a clean sample suitable for the automated measurement of narrow lines.

\section{\texorpdfstring{\naid}\ EW evolution analysis and discussion}
\label{sec:evolution}

The evolution of narrow lines from intervening material has been largely studied in the literature, and variations in the EW have been reported for various transients \citep[e.g.][]{Byrne23,deJaeger15}. Such observations have important implications for the presence and possible interaction with circumstellar matter or nearby patchy interstellar clouds. In high-resolution spectra of SNe~Ia, SN~2006X \citep{Patat07}, SN~2007le \citep{Simon09}, SN~2013gh \citep{Ferretti16}, and SN~2014J \citep{Graham15} showed changes in the EWs and profiles of the narrow intervening features. In the previous section, we showed with simulations that the EW measurement of narrow lines in high-resolution spectra is not affected by systematic biases; therefore, changes in profiles and intensities presented in these works probably represent true changes in the absorption of material in the line of sight of those SNe. On the other hand, we have shown that variations in low-resolution spectra should be analyzed with care. In particular, the effect of the varying morphology of the underlying ejecta P-Cygni profile must be considered. \cite{Wang19} found variable \naid\ absorption lines in low-resolution spectra of high-velocity SNe~Ia at early phases. We found a similar behaviour at similar phases in our low-resolution simulation: a decrease in the EW, followed by an increase, finally settling onto a constant evolution. As shown in Figure~\ref{fig:simEW-SNIa}, the changes in the EW are closely related to the slope of the continuum. When a continuum cut is considered, the variability in most objects disappears. This does not mean that the variability of the narrow absorption lines in low-resolution spectra cannot be real, but we suggest that careful cuts must be considered before performing such analysis and drawing conclusions. 

After including the cuts described in previous sections (see summary in Table~\ref{tab:cuts}), we evaluate the behaviour of the strength of the narrow \naid\ absorption lines as a function of time. Since EW measurements from an individual spectrum can be unreliable (Section~\ref{sec:nspec}), we require at least two spectra per epoch range for a given SN; we then stack them and recalculate the EW (see Section~\ref{sec:multmeas}). For this analysis, we explore various possibilities, using at least two or at least three spectra per SN in intervals of 10 to 20 days. While stacking SN spectra in time intervals will wash out shorter timescale variations, it ensures that we can study the time evolution of the EW with our methodology and low-resolution spectra. Once the spectra are stacked, we look at the evolution of the narrow lines both statistically and individually. Particularly, we focus on the EW of the line because it is more robust than the velocity (see section~\ref{sec:vel_meas}).

\subsection{Statistical evolution of EW}

We study the main SN types, SNe~Ia, SNe~II, SESNe and SNe-int, for statistical signs of evolution. Finding population-wide signs of evolution could give us important insight into the progenitor systems and nearby material of a given population. For instance, a large number of SNe~II are known to be subject to very early interaction with CSM \citep[e.g.][]{Bostroem23, Martinez23} 
that could perhaps be revealed in the narrow absorption line evolution. SNe-int are by definition subject to interaction with nearby material and thus good candidates to observe evolution in the narrow absorption lines, as is the case for SN~2011A\citep{deJaeger15}; however, the line is often absent despite strong CSM interaction\citep[e.g.][]{Ofek10, Pastorello18}. To investigate this, we require a common reference epoch for each group. We use the maximum $B$-band dates obtained through multi-wavelength light-curve fits with spectrophotometric templates of each SN type (see Appendix~\ref{ap:maxdates}). The epochs are normalized to this reference epoch and put in rest-frame. All spectra of a given SN in time intervals of, e.g. 10 days, are stacked, and the EW is measured. We investigate the evolution of the $\sigma$-deviations per epoch interval $i$ with respect to the all-epoch stacked EW measurement of each SN, i.e. $\overline{\mathrm{EW}}$, so that: 

\begin{equation}
\delta\sigma_i = \frac{\mathrm{EW}_{i}-\overline{\mathrm{EW}}}{\sqrt{\sigma_{\mathrm{EW}_i}^2+\sigma_{\overline{\mathrm{EW}}}^2}}. 
\end{equation}

\begin{figure}
\centering
\includegraphics[width=1.0\columnwidth]{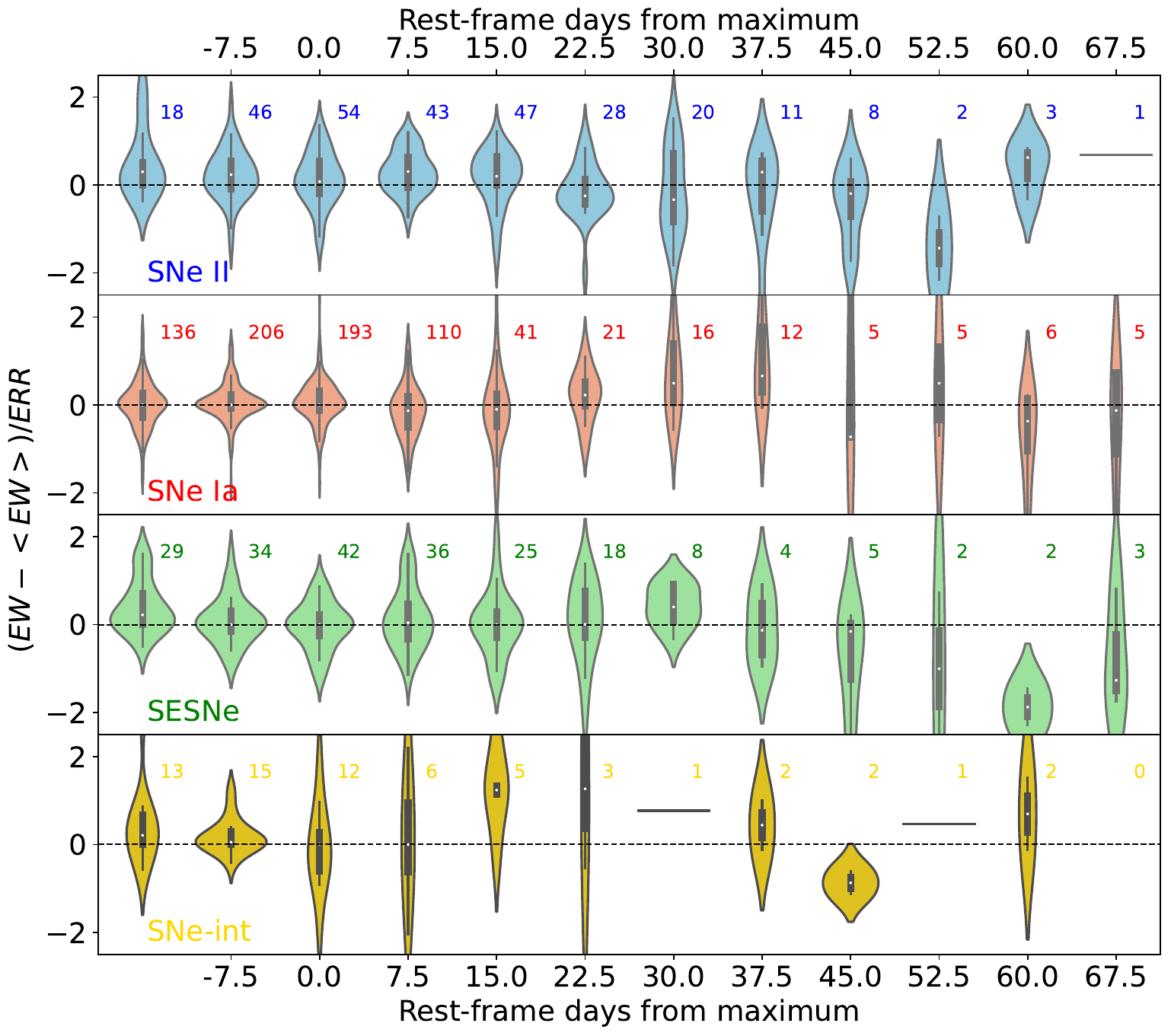}
\caption{\naid\ Violin plots of \naid EW evolution for each SN type (from top to bottom: SNe~II, SNe~Ia, SESNe and SNe-int) showing the mean and extreme of the distribution (white point in the box and edges of central vertical lines) and a kernel density probability estimation (coloured area). The number of SNe within each distribution at each epoch is indicated above each bin. We require at least two spectra per SN in intervals of 15 days for stacked EW measurements. 
}
\label{fig:evolnaidn2}
\end{figure}

Figure~\ref{fig:evolnaidn2} shows the evolution of the distribution of these deviations for the \naid lines of the four SN types considered. In general, we can see that there is no large evolution of these groups within the uncertainties. For SESNe, there seems to be a decrease after 50 days post maximum, although this is likely due to low-number statistics. Similarly, SNe~Ia appear to increase within 30 and 50 days post maximum, but with a small number of SNe at these epochs, this is mainly driven by one SN (SN~2012dn), as we will see in the next section. SNe-int also have very low statistics making any conclusion rather difficult. When investigating the evolution with different epoch intervals between 10-15 days and also requiring a minimum of three spectra per SN per interval instead of two, we obtain similar results. 

Additionally, we also investigate narrow lines from other species, such as \caii and \ki finding no evidence of evolution. However, it is possible that smaller sub-groups might show signs of changes in time. Given the small statistics after all our cuts, we leave this for a future investigation. In the next section, we analyse the evolution of individual well-sampled SNe. 

\subsection{Individual evolution of EW}

To analyze the evolution of the EW of narrow absorption lines for individual objects, we take the stacked EW of a minimum of two spectra in intervals of 10 days, as in the previous section. Besides a visual inspection, we apply two methods to identify possible candidates for evolution automatically:

\begin{itemize}
\item \textbf{Linear slope:} we fit the EW of all intervals of each SN to a simple line with a Bayesian fit from which the posterior distribution gives us the probability for the slope to be different from zero \citep{Kelly07}. A significantly non-zero slope ($p>95\%$) indicates that there is a general monotonic trend in the evolution of the SN. 
\item \textbf{Cumulative sum:} to try to capture more complicated evolution, we also do a cumulative sum analysis \citep[{\sc CUSUM},][]{Page54-cusum} that monitors deviations from a target, in this case, the EW value obtained from the stack of all the spectra (at all epochs) of a given SN, $\overline{\mathrm{EW}}$. Then the algorithm measures accumulated deviations, chosen here as a drift of half the uncertainty on the stacked EW,  $0.5\sigma_{\overline{\mathrm{EW}}}$, to both positive and negative directions, so that:

\begin{eqnarray*}
H_0 &=& 0 \\
L_0 &=& 0 \\
H_i &=& max\left[0,H_{i-1} + (\mathrm{EW}_i - \sigma_{\mathrm{EW}_i}) - (\overline{\mathrm{EW}} + 0.5\sigma_{\overline{\mathrm{EW}}})\right] \\
L_i &=& min\left[0,L_{i-1} + (\mathrm{EW}_i +\sigma_{\mathrm{EW}_i})  - (\overline{\mathrm{EW}} - 0.5\sigma_{\overline{\mathrm{EW}}})\right]\, \\
\end{eqnarray*}

where $H_i$ and $L_i$ indicate the positive and negative cumulative deviations updated at each time step $i$. We also take into account the uncertainty in each individual EW$_i$. This measures the cumulative deviations starting from the earliest epoch interval of the SN and increasing in time. In order to not give higher weight to the initial time bins, we also start the algorithm from the latest epoch interval going in decreasing time steps. The final cumulative $H$ and $L$ are the average of both sequential cumulative deviations. SNe with more than two cumulative deviations above the error on the stacked EW are considered viable candidates.

\end{itemize}

\begin{figure*}
\centering
\includegraphics[width=0.49\textwidth]
{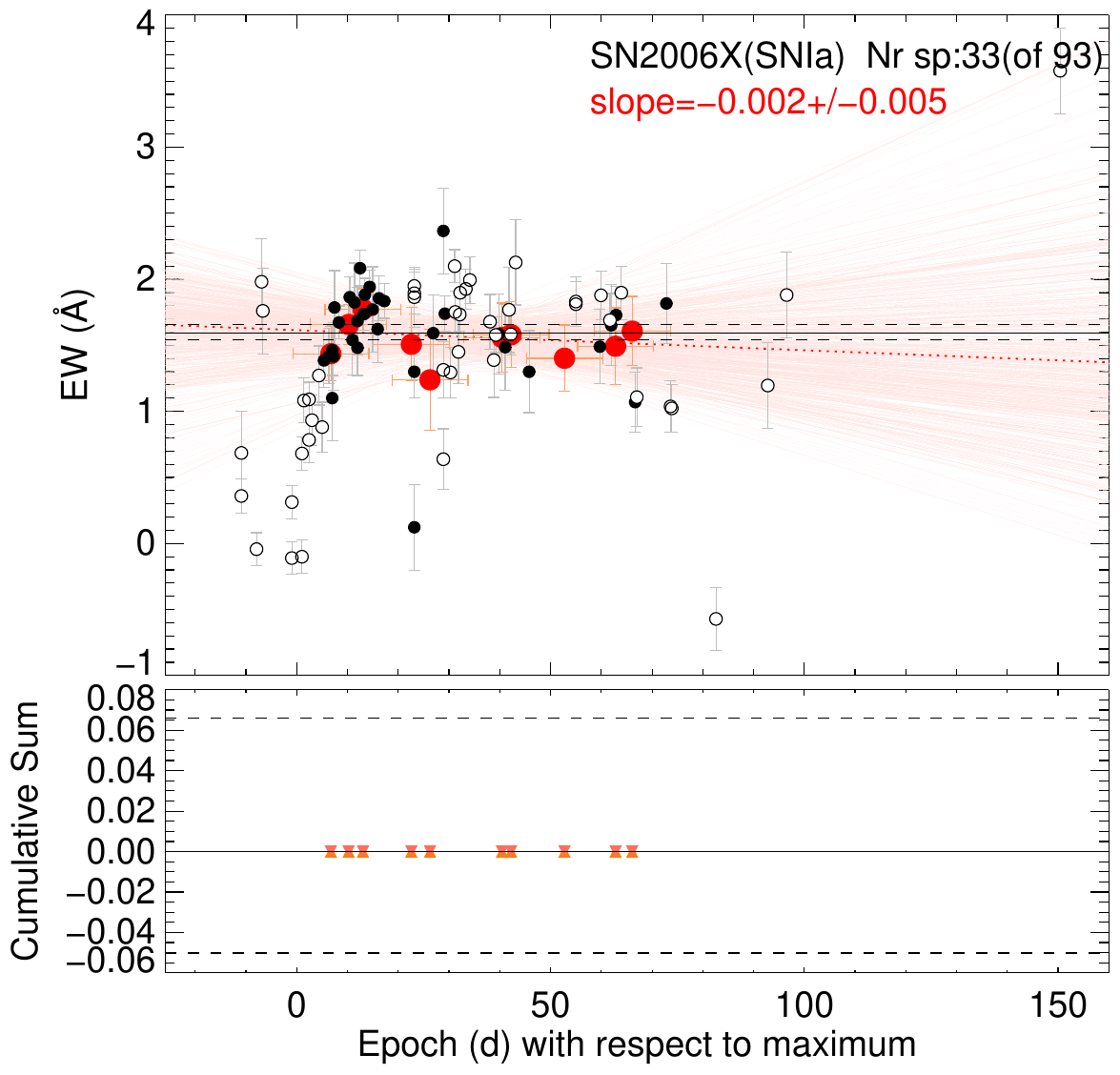}
\includegraphics[width=0.49\textwidth]
{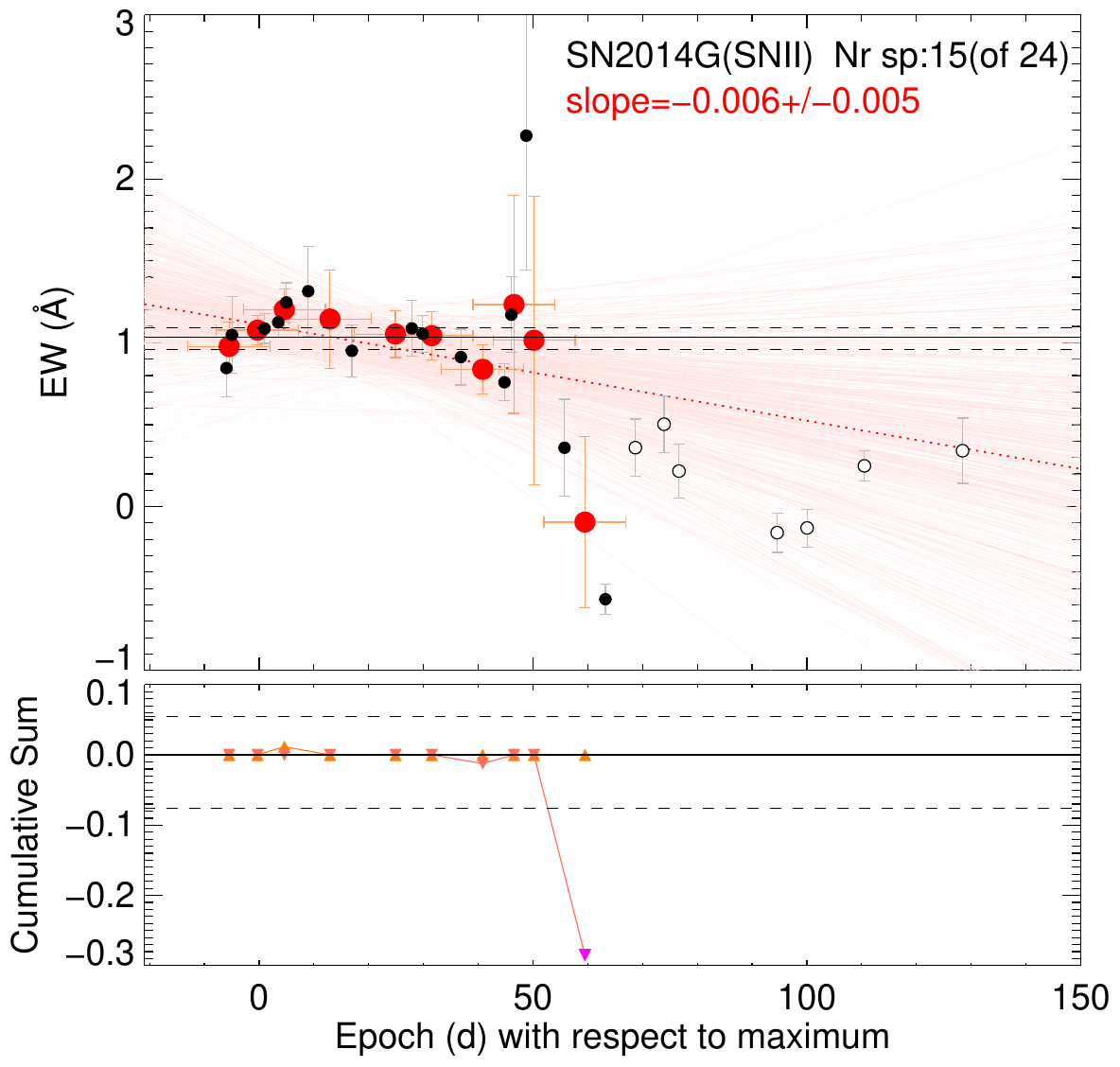}
\caption{\textbf{Top}: \naid\ EW evolution for SN~2006X  (left) and SN~2014G (right) before/after cuts (empty/filled black points) and stacked measurements of points after culls in intervals of 15 days requiring minimum 2 spectra (red points). Horizontal lines indicate the stacked EW and error from all spectra that passed the cuts. A Bayesian linear fit to the red points is shown with light red lines (1000 realizations) and a corresponding median with a dotted red line. The resulting slope is negative with 68.7\% and 93.7\% probability, respectively for both SNe. \textbf{Bottom}: Cumulative deviation analysis above/below (upper/lower triangles) the stacked EW (corresponding to the zero horizontal line) and its associated errors (dashed lines). Significant points are shown in purple.
}
\label{fig:99em-evolanal}
\end{figure*}

Figure~\ref{fig:99em-evolanal} shows these two methods applied to the measurements of SN~2006X (SN~Ia) and SN~2014G (SN~II) after the cuts (filled black points). Doing a Bayesian linear fit to the stacked EWs per epoch interval (red points) results in a slope consistent with zero for SN~2006X and no significant points in the cumulative deviation analysis. SN~2014G, on the other hand, has a slope with a probability of 93.7\% to be non-zero and one significant cumulative deviation point. Both of these SNe are considered to be not evolving in time according to our diagnostics. However, it is important to observe that if we were to do our analysis on all data points without cuts (empty black points), these SNe would have non-zero slope probabilities of 88.4\% and 100\%, with  4 and 12 significant cumulative deviation points, respectively. Both SNe would be considered to have an evolution. This shows the importance of our cuts, particularly the continuum slope that prevents strong blending with the broad P-Cygni profile (see Section~\ref{sec:broadline}).

When analysing various narrow lines for all SNe after cuts and stacking at least two spectra per 10 or 15-day intervals, we find that very few SNe are tagged as possibly evolving according to our two diagnostics. For every case, we ensure that no equivalent evolution is seen for the same lines of the MW and inspect them visually. The few SNe tagged (e.g. SN~2001N, SN~2012dn) have few points in the nebular phase with stronger EW. However, these are a very small number and, in some cases, still noisy after stacking per interval. In the case of SN~2012dn, a 03fg-like "super-Chandra" SN~Ia \citep{Chakradhari14, Taubenberger19}, the narrow \naid at $\sim40$ days after the maximum gets mixed with features due to iron-group elements and it is difficult to disentangle the two. 

SNe known to show variation in their lines with high-resolution spectra, e.g. SN~2006X, SN~2007le, SN~2013gh or SN~2014J, are not picked up by our culls, but we have predominantly low-resolution spectra for which such small variations remain within the uncertainties. We have no access to high-resolution spectra for SN~2006X, SN~2013gh and only one for SN~2007le. For SN~2014J, if we restrict our analysis to the 10 high-resolution spectra we have and do not make any cuts on the continuum slope, we find no evolution of the EW for \naid\ and no evolution of the three spectra that cover the wavelengths of \ki. \citealt{Graham15} found evolution in \ki lines in other spectra of even higher resolution (R$>$100000). In the case of the type IIn SN~2011A \citep{deJaeger15}, we do see the increase in EW with time that they report. Nonetheless, with only 5 spectra after cuts (and only two stacked points per 15-day interval), the linear slope is not significantly positive, and the stacked EW of all spectra is too influenced by the late spectra with increasing EW, such that no significant deviation points are found.

We conclude that the current dataset, after all our cuts and with our methodology, is consistent with no variation in the EW of narrow lines over time within the uncertainties. More subtle variations are naturally possible but necessitate better sampling and, foremost, higher spectral resolution.

\section{Conclusions}
\label{sec:conclusions}

In this paper, we have presented a statistical analysis of the evolution of narrow line-of-sight lines from intervening material (\naid, \caii\,H\&K, \ki, DIBs) visible in nearby SNe. We have analyzed over 11000 spectra of more than 1600 SNe with various resolutions. To quantify the properties of these narrow lines, we developed a robust automated methodology allowing for the estimation of their EWs and velocities. When searching for possible EW evolution, we found systematic variations in the \naid\ line from the MW and the host galaxies. Changes in the \naid\ depth from the MW narrow interstellar lines are in principle not expected because the material in our own Galaxy is at sufficiently large distances from the SN. 
The strength of the narrow absorption lines in low and mid-resolution spectra may thus be affected by different external factors that we explore: the signal-to-noise of the spectra, the spectral resolution, the slit size and orientation, the number of spectra used and most importantly, the interference of the broad P-Cygni lines from the SN. We found that these biases, particularly those associated with presence of the broad lines, induce a conspicuous apparent evolution (variability) over time in many SNe. For instance, in SNe~II, the narrow sodium line overlaps with the broad sodium from the SN ejecta, and in SNe~Ia the narrow sodium overlaps with the P-Cygni profile of \sii. Indeed, we confirm this with simulations in which we input fake narrow lines into different locations of the broad SN P-Cygni lines and calculate the recovered EW. In our simulations of low-resolution spectra, we found large significant changes in the EW depending on the underlying profile. However, in our high-resolution simulations, the strength of the narrow lines was not affected by the interference of the broad SN P-Cygni profiles.    

Based on these findings, we presented an easy way to detect and exclude cases of high contamination of the SN profile in the narrow line: when the normalized slope of the continuum is too high, i.e. $m_c>0.002$\,\AA$^{-1}$, there is a broad component underneath blending with the narrow line. We show that excluding these cases with a steep continuum is essential in the study of narrow lines from intervening material in the line of sight. Not taking this into account may lead to important biases in the frequently used estimations of dust extinction through the EW$\propto A_V$ relation (Eq.~\ref{eq:EW-Av}) in low-resolution spectra, and it may lead to wrong inferences of the evolution of the lines and the presence of nearby material. Previous claims of the evolution of narrow lines in low-resolution spectra should be revisited in light of this study. 

We also strongly encourage the use of more than one low-resolution spectrum in the estimation of the EW. Fluctuations of the S/N, the spectral resolution, and the morphology of the underlying P-Cygni profile over time can generate changes in these lines that are washed out with a larger number of spectra stacked to obtain a more accurate EW. We show that at least 2-3 spectra are needed to measure robust EWs with conservative uncertainties.  %

Finally, after considering all these effects, we analysed the temporal evolution of line-of-sight features in a large sample of nearby SNe to detect any possible statistical variation in their EWs. We found no overall evolution for three principal SN groups: SNe~II, SNe~Ia and SESNe, in all narrow lines considered. Significant individual SN variations were also ruled out based on our methodology and low-resolution spectra.

In this work, we have paved the way for robust measurement of narrow absorption lines in a variety of spectra from different instruments and resolutions. Armed with this methodology, we can now use a large sample to study the variations of the SN-averaged narrow lines according to SN type, to SN properties, and to host galaxy characteristics. A wealth of information on the pervading intervening material is contained in those lines providing essential clues to the immediate and farther vicinity of SNe and their progenitors, with far-reaching consequences for our understanding of the ISM cycle of galaxies and the persistent question of extinction in supernova cosmology. 

\begin{acknowledgements}

We sincerely thank the referee for very useful comments that improved the quality of the paper. We heartily thank all the research groups and observers for making their data public. This research has made use of the Berkeley SNDB database as well as the CfA Supernova Archive, which is funded in part by the National Science Foundation through grant AST 0907903. 
S.G.G. acknowledges the financial support from the visitor and mobility program of the Finnish Centre for Astronomy with ESO (FINCA), funded by the Academy of Finland grant nr 306531. S.G.G. and A.M.M. acknowledge support by FCT under Project CRISP PTDC/FIS-AST-31546/2017 and Project~No.~UIDB/00099/2020.
C.P.G. acknowledges financial support from the Secretary of Universities and Research (Government of Catalonia) and by the Horizon 2020 Research and Innovation Programme of the European Union under the Marie Sk\l{}odowska-Curie and the Beatriu de Pin\'os 2021 BP 00168 programme. C.P.G. and L.G. recognize the support from the Spanish Ministerio de Ciencia e Innovaci\'on (MCIN) and the Agencia Estatal de Investigaci\'on (AEI) 10.13039/501100011033 under the PID2020-115253GA-I00 HOSTFLOWS project, from Centro
Superior de Investigaciones Cient\'ificas (CSIC) under the PIE project 20215AT016, and the program Unidad de Excelencia Mar\'ia de Maeztu CEX2020-001058-M. L.G. also thanks the financial support from the European Social Fund (ESF) ”Investing in your future” under the 2019 Ram\'on y Cajal program RYC2019-027683-I. J.P.A. acknowledges support by ANID, Millennium Science Initiative, ICN12\_009. A.M.G. acknowledges financial support by the European Union under the 2014-2020 ERDF Operational Programme and by the Department of Economic Transformation, Industry, Knowledge, and Universities of the Regional Government of Andalusia through the FEDER-UCA18-107404 grant. S.M. acknowledges support from the Research Council of Finland project 350458.\\

\end{acknowledgements}

%
\bibliographystyle{aa} 
\bibliography{Bibliography} 
%


\appendix

\section{Wavelength calibration uncertainty}\label{ap:wavecalib}

\begin{figure}
\centering
\includegraphics[width=0.49\textwidth]{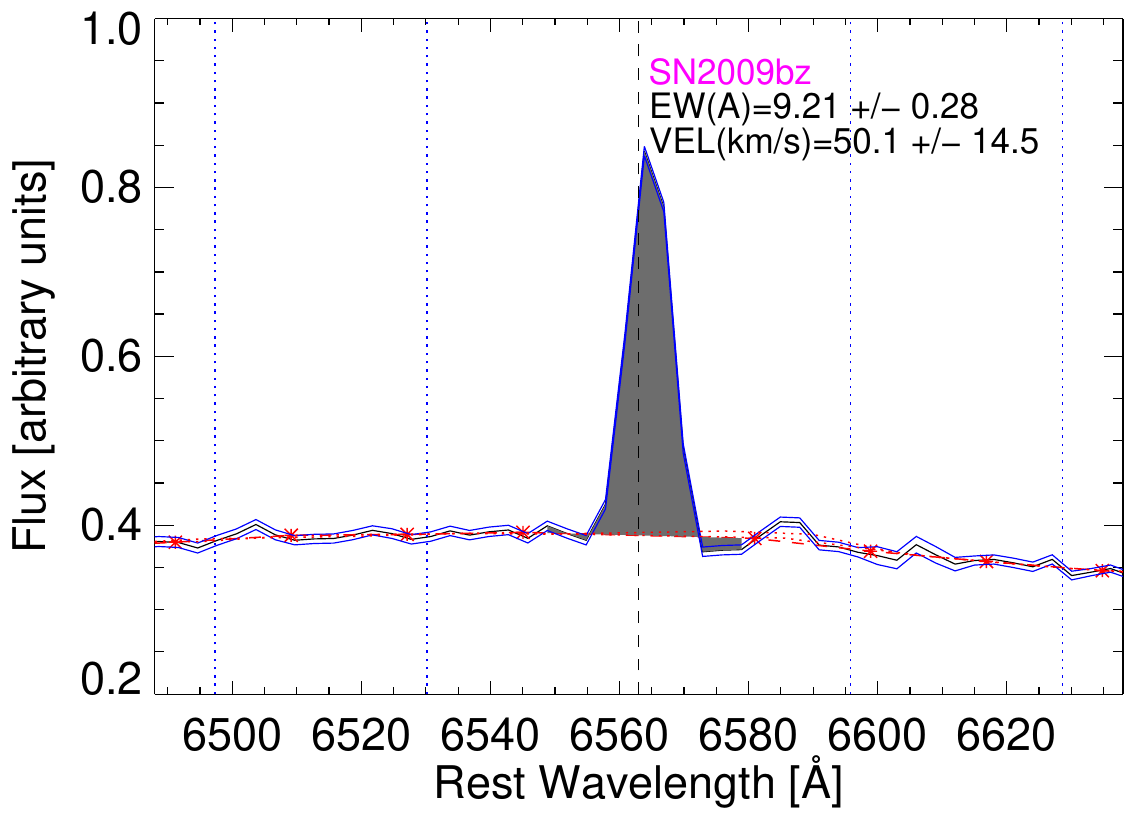}
\caption{Spectrum of SN~2009bz around the emission line H$\alpha$ and automated measurement of EW and velocity. The quoted uncertainty in velocity does not include the wavelength calibration. The dispersion of the velocities of the three lines (H$\alpha$, H$\beta$ and [OIII]) for this spectrum is 33.2 km/s.}
\label{ap:09bz}
\end{figure}

The heterogeneous dataset used in our analysis has been wavelength-calibrated by different groups for different instruments and observing conditions. Here we estimate an average uncertainty of the wavelength calibration for the entire sample. For this, we randomly take 1556 spectra of 180 core-collapse SNe from our sample and use our automated technique (see~\S\ref{sec:vel_meas}) around three narrow emission lines characteristic of gas-rich environments: H$\alpha$ (6563\AA), H$\beta$ (4681\AA) and [OIII] (5007\AA). We use the following integration windows: $\pm$ 750 km/s, $\pm$ 750 km/s and $\pm$ 900 km/s, respectively. An example measurement of H$\alpha$ for a spectrum of SN~2009bz is shown in Figure~\ref{ap:09bz}. The dispersion of the three line velocities for this spectrum gives 33.2 km/s. Of the 1556 spectra, we select 696 that clearly show at least two of the three lines, i.e. requiring that EW/$\sigma_{\mathrm{EW}} > 1.3$ and EW $> 0.5$\AA; and we calculate the median dispersion for the sample finding 35.0 km/s. Doing the same analysis using Gaussian fits instead, we obtain proper fits with EW$>0.5$ in at least two lines for 767 spectra for which the average of the velocity dispersion gives 36.6 km/s. This is in remarkable agreement with the automated value confirming that this dispersion is not dependent on the discussed methodologies. This value could reveal calibration uncertainties or real physical differences in the elements of the gas clouds, or a combination of both. We conservatively take it here as an intrinsic calibration uncertainty that is added in quadrature to other uncertainties of the method.

\section{$B$-band maximum dates}
\label{ap:maxdates}

\begin{figure}
\centering
\includegraphics[width=0.48\textwidth]{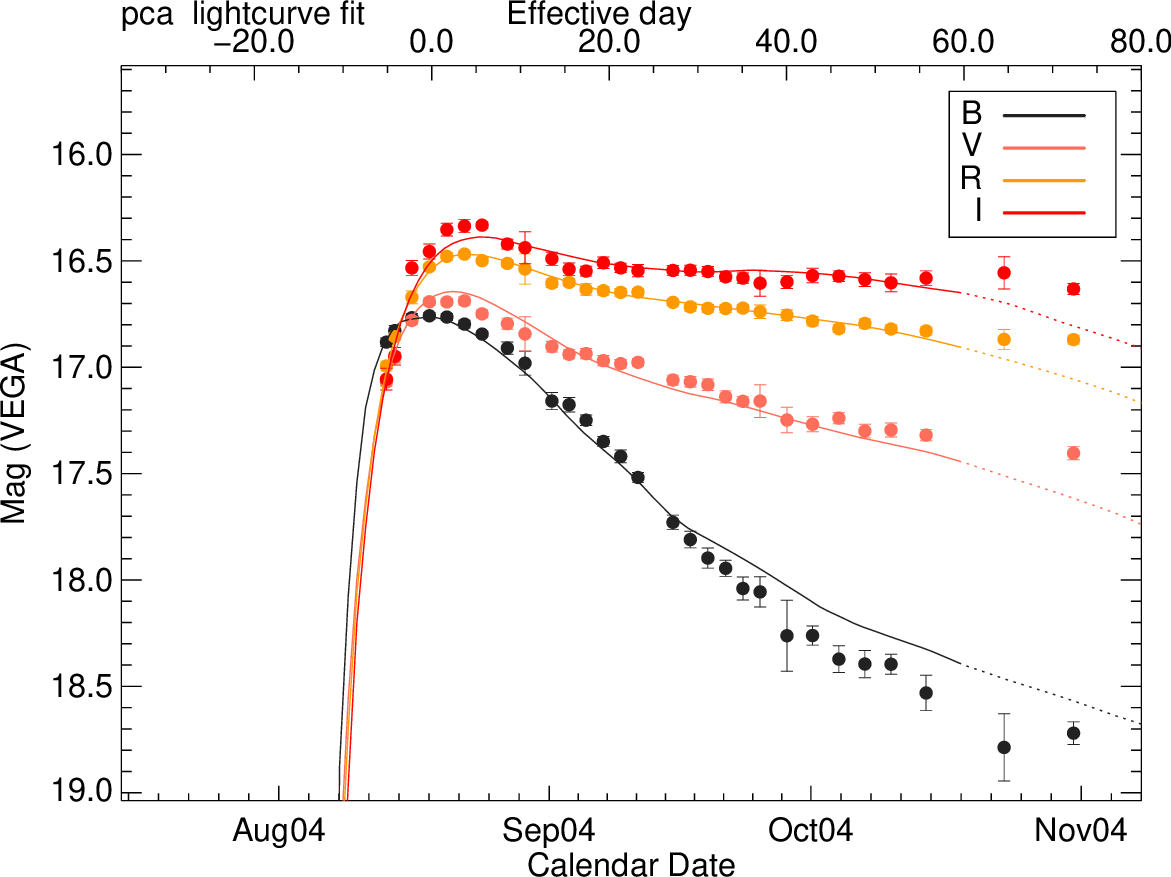}
\includegraphics[width=0.48\textwidth]{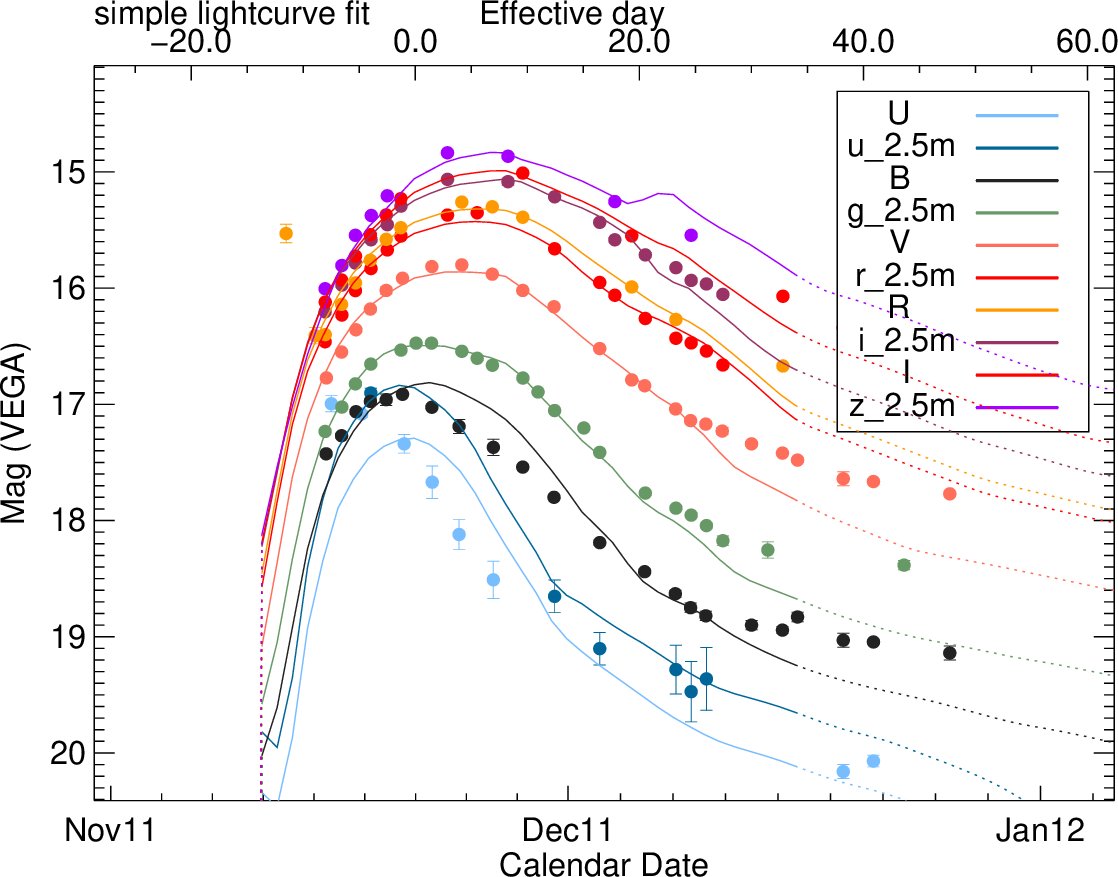}
\includegraphics[width=0.48\textwidth]{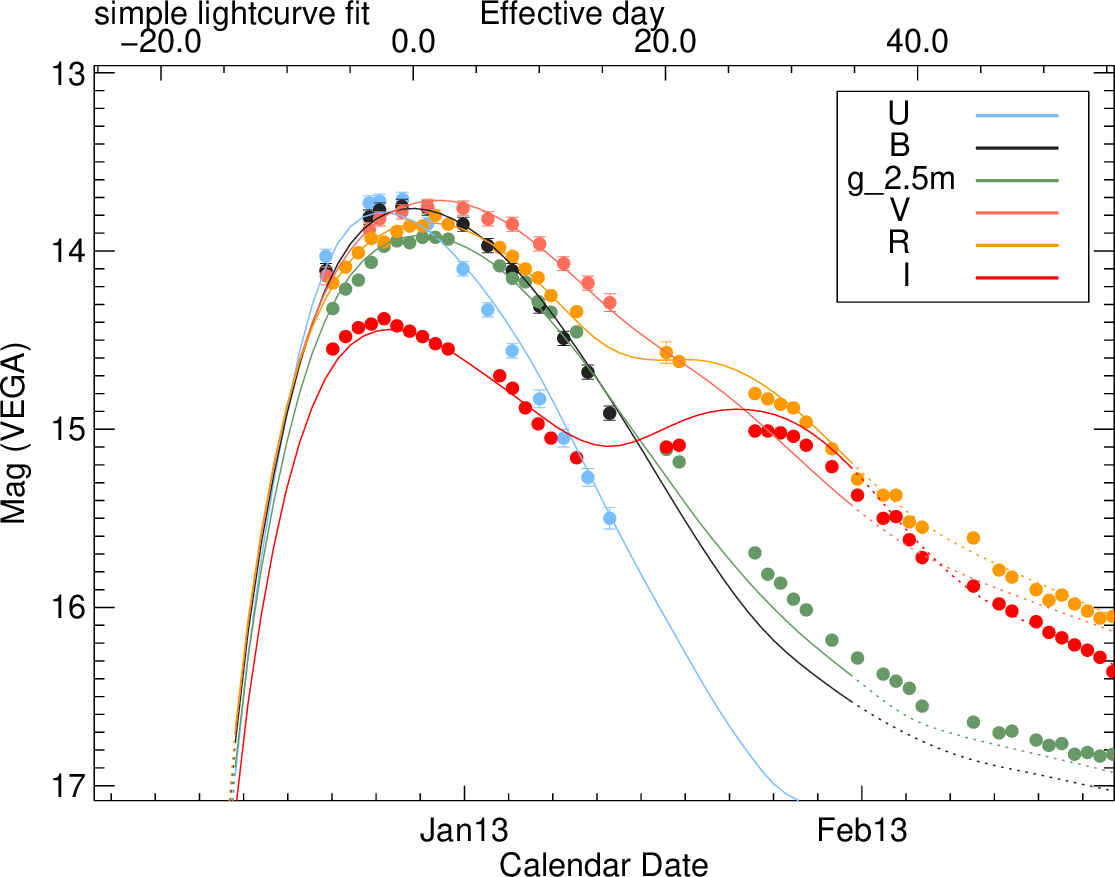}
\caption{Example of multi-band light-curve fits with SiFTO \citep{Conley08} and different spectrophotometric templates to SN2004du (SN~II, up), SN2011hs (SN~IIb, middle) and SN2012hr (SN~Ia, bottom).}
\label{fig:lcfits}
\end{figure}

To obtain the $B$-band maximum reference epochs used in section~\ref{sec:evolution}, we perform multi-band light-curve fits to every SN. We use a light-curve fitter commonly used for SNe~Ia \citep{Conley08} but with different spectrophotometric templates for each type: the template by \citet{Hsiao07} for SNe~Ia, the template of \citet{Nugent02} for SESNe and a model of SNe~II used in the PLAsTiCC challenge \citep{Kessler19} based on empirical spectral energy distributions that are a linear combination of three ‘eigenvector’ components. For SNe-int, we try both, the template for SNe~II and for SESNe, choosing visually the best obtained fit. Example fits are shown in Figure~\ref{fig:lcfits}. All fits are visually inspected, and SNe with poor data or bad fits are discarded. Although core-collapse SNe are not as homogeneous as SNe~Ia, and this fitter was devised for SNe~Ia, this simple approach gets rough estimates that allow us to have epochs in the same reference system. We also attempted to use the explosion epochs as a reference for SNe~II \citep{Gutierrez17a}, confirming no general evolution of EW with time in all three SN types.

\section{Figures}
\label{ap1}

\begin{figure*}
\centering
\includegraphics[width=\textwidth]{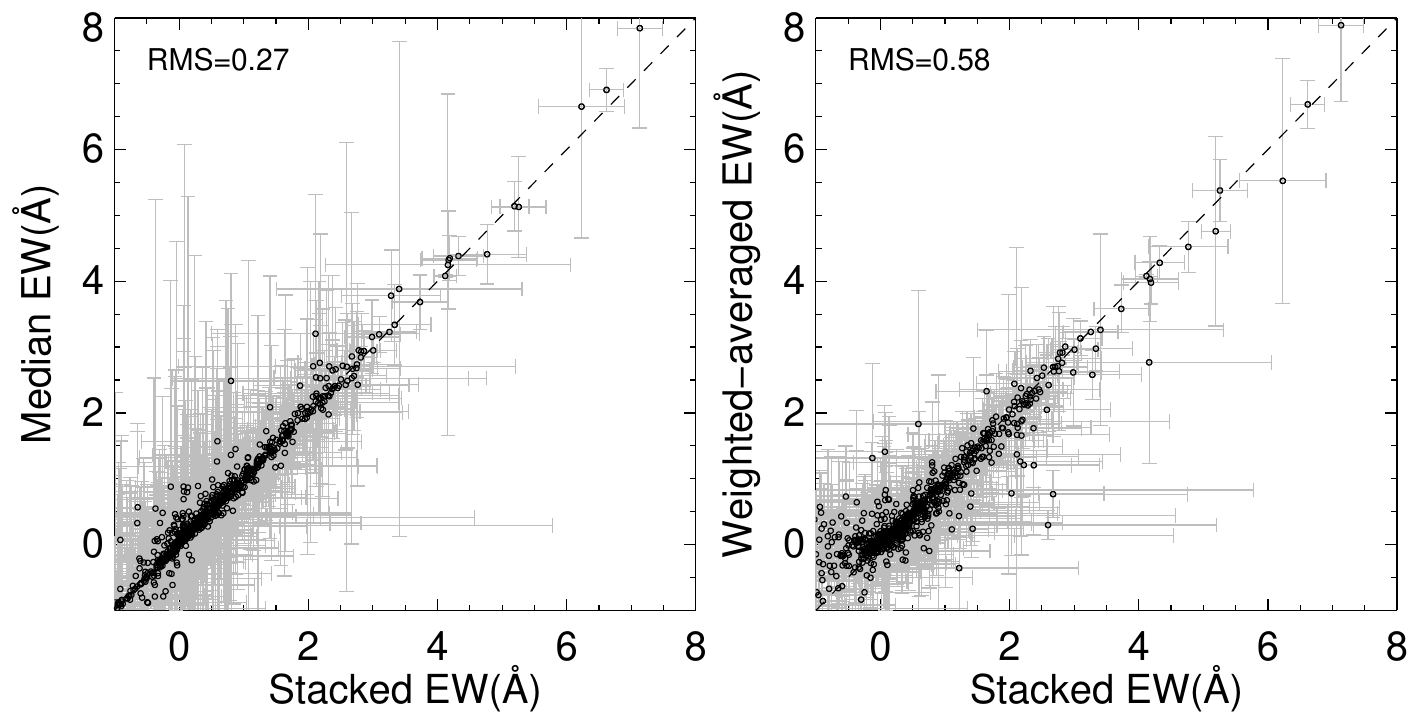}
\caption{Comparison of \naid EW measured from the stack of multiple spectra of each SN vs the median of individual EW (left) and the weighted average (right). The dashed lines indicate a $x=y$ relation.}
\label{fig:stackcomp}
\end{figure*}

\begin{figure*}
\centering
\includegraphics[width=1.0\textwidth]{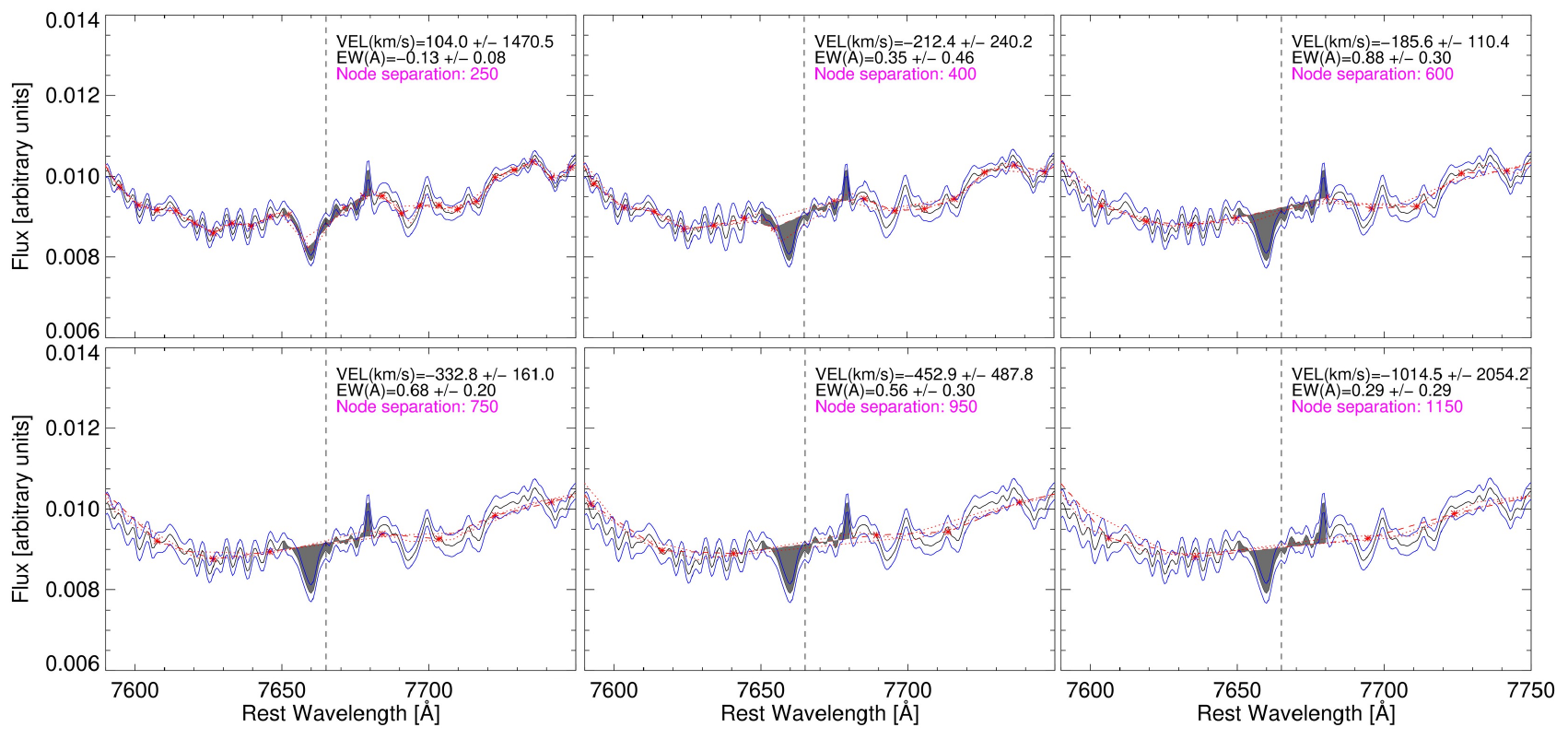}
\caption{SN~2014J spectrum around the \ki 7665\AA\, line showing different measured continua by increasing the node separation from top left to bottom right: 250, 400, 600, 750, 950 and 1150 km/s. The black line is the spectrum, the blue lines are the estimated flux errors, the red stars are the nodes and the red dashed line is the continuum crossing the nodes. Red dotted lines represent the continua at $\pm25\%$ of the node separation being considered. The grey-shaded area is the integrated line within $\pm$600 km/s. The dashed vertical line is the central wavelength of the line. }
\label{fig:cont}
\end{figure*}

\begin{figure*}
\centering
\includegraphics[width=1.0\textwidth]{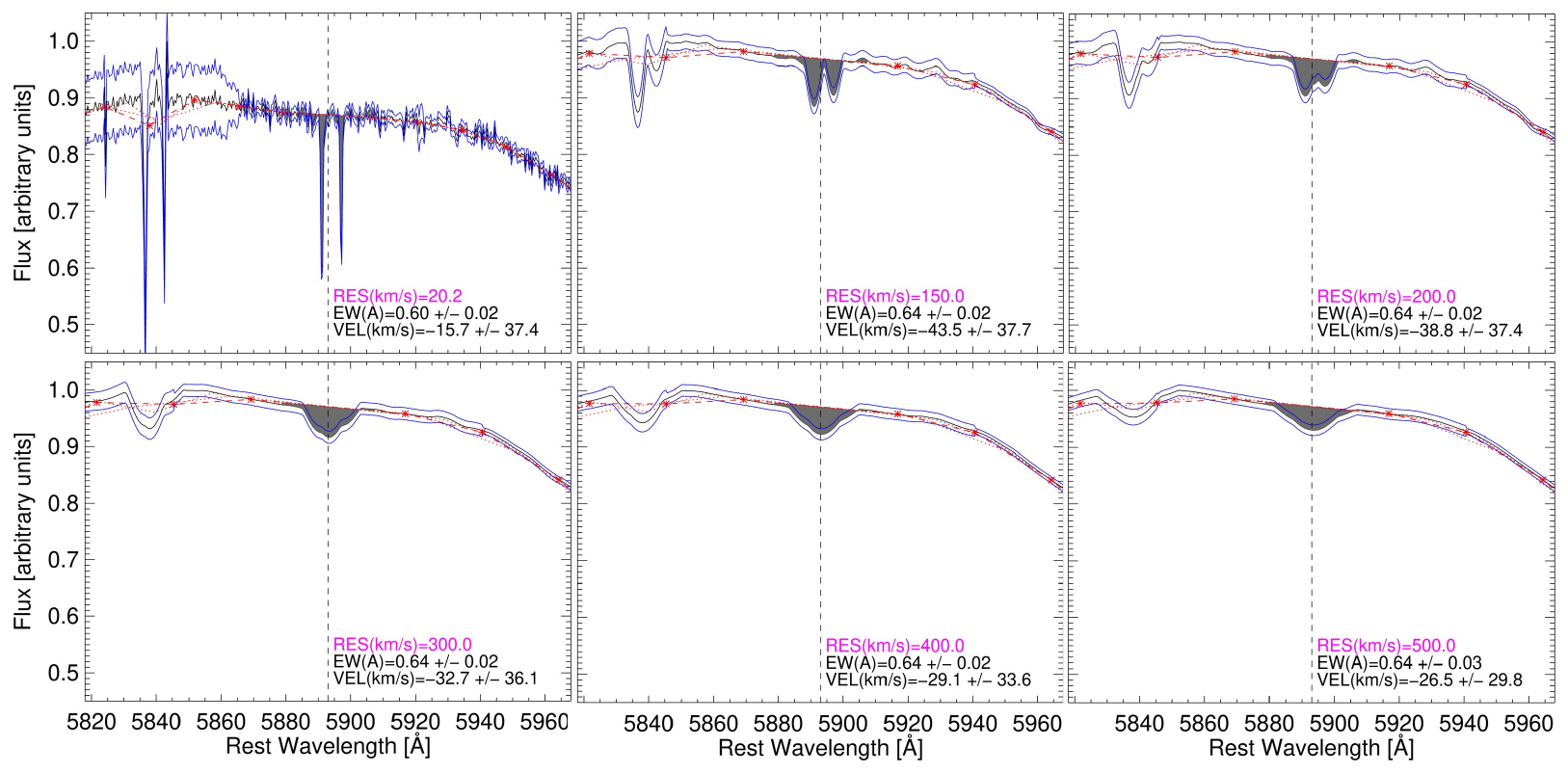}
\caption{SN~2010ev high-resolution spectrum (top left) smoothed with decreasing kernel resolutions (increasing $\Delta \mathrm{v}_{\mathrm{res}}=100-500$km/s) around \naid  from left to right and top to bottom. The estimated error on the flux is shown in blue. The automated methodology outlined in section~\ref{sec:measurements} calculates a continuum (shown in red) and integrates through a fixed spectral window (grey shading).}
\label{fig:specres_sim}
\end{figure*}

\begin{figure*}
\centering
\includegraphics[width=1.0\textwidth]{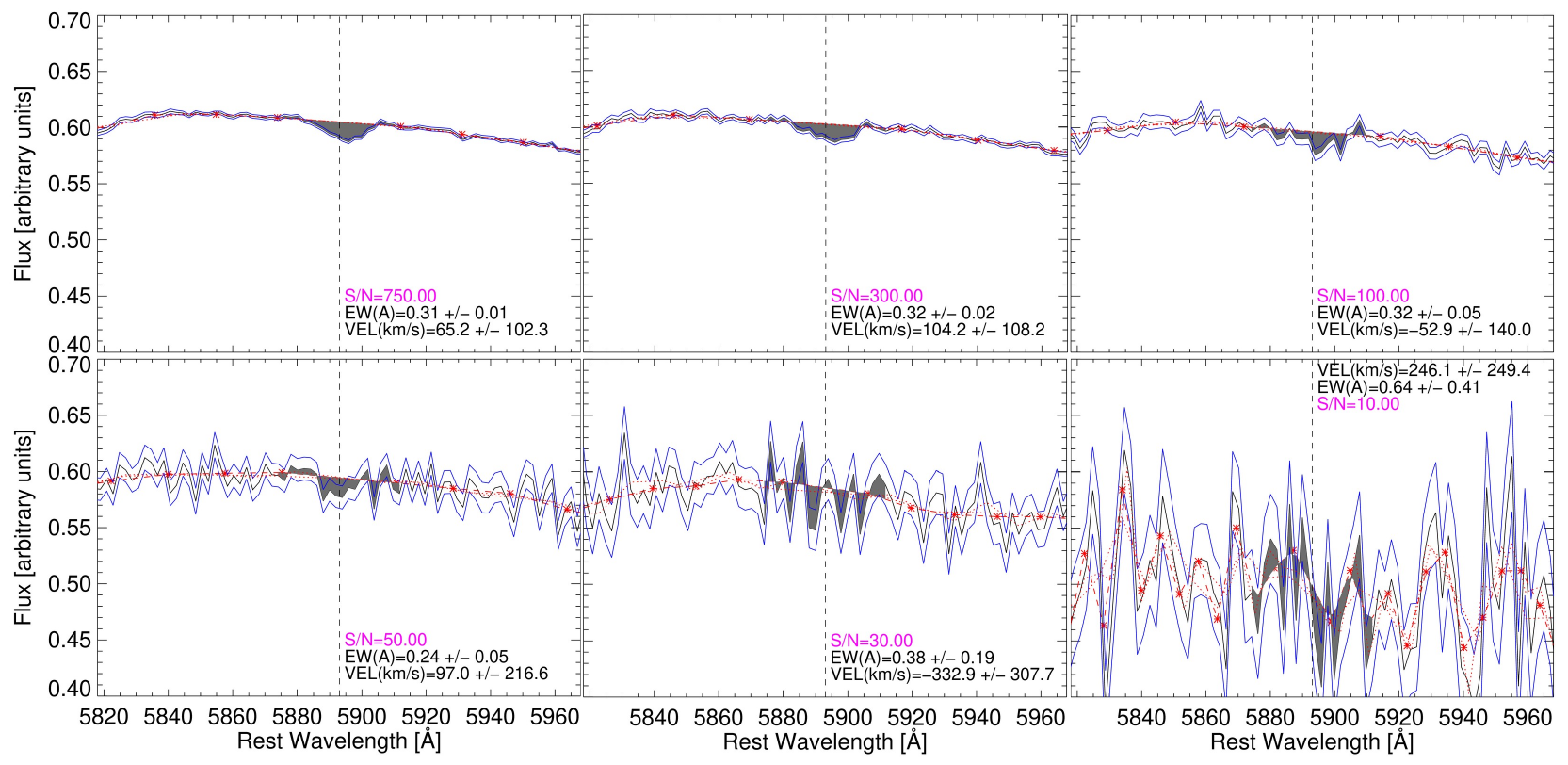}
\caption{SN~2002an high-signal spectrum (top left) perturbed to various decreasing S/N around \naid from left to right and top to bottom. The estimated error on the flux is shown in blue. The automated methodology outlined in section~\ref{sec:measurements} calculates a continuum (shown in red) and integrates through a fixed spectral window (grey shading). The vertical dashed line shows the rest-frame central line of \naid.}
\label{fig:signois_sim}
\end{figure*}

\begin{figure*}
\centering
\includegraphics[width=0.495\textwidth]{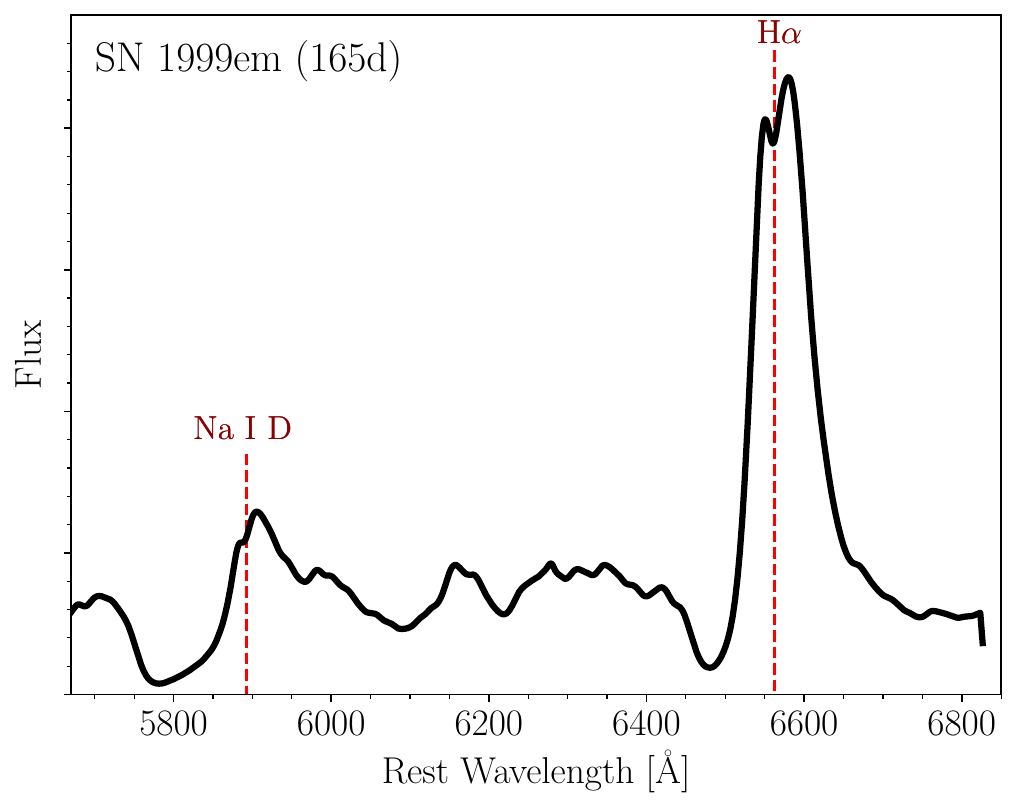}
\includegraphics[width=0.495\textwidth]{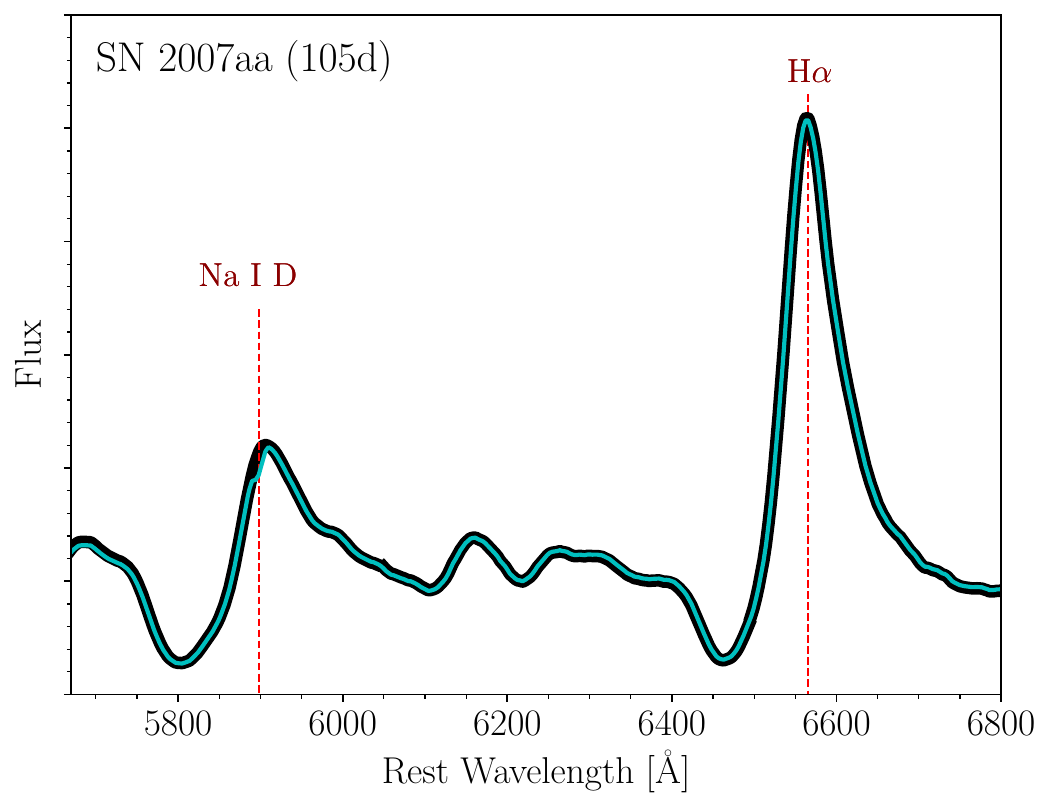}
\caption{Observed spectra (black) of SN~1999em \textbf{(left)} and 2007aa \textbf{(right)} at 165 and 105 days from the explosion. The vertical dashed lines show the rest-frame position of the strongest lines. The simulated spectrum of SN~2007aa, where a fake \naid\ narrow line was added at $\lambda=5893$ \AA\ (EW$=1.1$ \AA), is presented on top of the observed spectrum (cyan). The peak of the \naid seems redshifted in SN~1999em, compared to H$\alpha$, which is at the rest-position. A comparable shift can be obtained when a narrow line is located on the top of the broad SN P-Cygni line of SN~2007aa.}
\label{fig:sneII}
\end{figure*}

We present here additional figures: Figure~\ref{fig:stackcomp} shows the comparison of the EW measured from the stacked flux-to-continuum ratios of several spectra of a given SN (see~\S\ref{sec:multmeas}) versus the median and the weighted average of individual spectra for a given SN (the uncertainties come from the median absolute deviation and the weighted error, respectively); Figure~\ref{fig:cont} displays the impact of the continuum node separation on the EW measurement (see~\S\ref{sec:ew_meas}); Figures~\ref{fig:specres_sim} and \ref{fig:signois_sim} shows how degrading the spectral resolution (\S\ref{sec:specres}) and the signal-to-noise (\S\ref{sec:signois}) affect the narrow line and its EW measurement; and Figure~\ref{fig:sneII} demonstrates that the narrow line at nebular times can emulate a blueshift in the line from the SN ejecta (see~\S\ref{sec:broadline}).

\section{Tables}
\label{ap2}

In table~\ref{table_SNe} we present all SNe used in this analysis with their type, redshift, number of spectra and the references for their spectra.

\renewcommand{\thetable}{A\arabic{table}}
\setcounter{table}{0}
\clearpage
{\onecolumn
\centering
\setlength{\tabcolsep}{2pt}
\scriptsize
\begin{longtable}[t!]{lcccl}
\caption{Sample of SNe initially included in our sample.}
\label{table_SNe} \\
\hline
SN                & Type     &  Redshift & \# of spectra &  References                                              \\	
\hline
                 &               & SNe Ia    &      &                                                                \\
\hline
       SN~1937C  &           Ia  &    0.001  &   2  &                                           \citealt{Casebeer00} \\
       SN~1937D  &           Ia  &    0.002  &   1  &                                           \citealt{Casebeer00} \\
       SN~1939A  &           Ia  &    0.003  &   2  &                                           \citealt{Casebeer00} \\
       SN~1954B  &           Ia  &    0.005  &  10  &                                           \citealt{Casebeer00} \\
       SN~1956A  &           Ia  &    0.004  &   8  &                                           \citealt{Casebeer00} \\
\hline
\hline
\end{longtable}
Additional references: [1] \citealt{Folatelli13}; [2] \citealt{Silverman12}; [3] \citealt{Blond12}; [4] \citealt{Gutierrez17a}; [5] \citealt{Stahl20};  [6] \citealt{Modjaz14}; [7] \citealt{Shivvers19};  [8] PESSTO data release: \citealt{Smartt15a}; [9] Publically available from WISEREP: \citealt{Yaron12}. 
\clearpage
}


\label{lastpage}

\end{document}